\documentclass[12pt]{article}
\usepackage[dvipdfmx]{graphicx}
\usepackage{epsfig}
\usepackage{amsfonts}
\usepackage{amsmath,amssymb}
\usepackage[dvipdfmx]{color}
\usepackage{fancybox} 
\usepackage{ulem}
\numberwithin{equation}{section}

\baselineskip=\normalbaselineskip

\eject  

\newcommand{\Bra}[1]{\left\langle\, #1\,\right|}
\newcommand{\Ket}[1]{\left|\, #1\,\right\rangle}
\newcommand{\Bracket}[2]{\left\langle\, #1\,|\, #2\,\right\rangle}

\def\mat#1{\matt[#1]}
\def\matt[#1,#2,#3,#4]{\left(%
\begin{array}{cc} #1 & #2 \\ #3 & #4 \end{array} \right)}
\def\vect#1{\vectt[#1]}
\def\vectt[#1,#2]{\left(%
\begin{array}{c} #1 \\ #2 \end{array} \right)}

\def\natural{\mathbb{N}}
\def\real{\mathbb{R}}
\def\complex{\mathbb{C}}

\def\ha{\hat{a}}
\def\had{\hat{a}^\dagger}

\def\p{\partial}
\def\hx{\hat{x}}

\def\bea#1\ena{\begin{align}#1\end{align}}
\def\nn{\nonumber\\}

\def\nn{\nonumber\\}

\allowdisplaybreaks
\begin{document}
\begin{titlepage}

\setcounter{page}{0}
\renewcommand{\thefootnote}{\fnsymbol{footnote}}

\begin{flushright}
	{\small UTHEP-715} \\
\end{flushright}

\begin{center}
{\large\bf
Commutative Geometry for Non-commutative D-branes\\ by Tachyon Condensation\\
}

\vspace{5mm}
Tsuguhiko Asakawa$^{1}$,%
\footnote{\tt asakawa@maebashi-it.ac.jp}
Goro Ishiki$^{2,3}$,%
\footnote{\tt ishiki@het.ph.tsukuba.ac.jp}
Takaki Matsumoto$^{3}$,%
\footnote{\tt  matsumoto@het.ph.tsukuba.ac.jp}\\
So Matsuura$^{4}$,%
\footnote{\tt s.matsu@phys-h.keio.ac.jp}
and
Hisayoshi Muraki$^{5}$%
\footnote{\tt hmuraki@sogang.ac.kr}

\vspace{5mm}

{\small{\it
$^{1}$ Department of Integrated Design Engineering,\\
Maebashi Institute of Technology,
Maebashi, 371-0816, Japan \\
$^{2}$ Tomonaga Center for the History of the Universe, \\
University of Tsukuba, Tsukuba, Ibaraki 305-8571, Japan \\
$^{3}$ Graduate School of Pure and Applied Sciences, \\
University of Tsukuba, Tsukuba, Ibaraki 305-8571, Japan \\
$^{4}$  Department of Physics, Hiyoshi Campus, \\ 
and Research and Education Center for Natural Science, \\
Keio University, 4-1-1 Hiyoshi, Yokohama, 223-8521, Japan \\
$^{5}$ Department of Physics, Sogang University, Seoul 04107, Korea 
}}
\end{center}


\begin{abstract}
There is a difficulty in defining the positions of the D-branes
when the scalar fields on them are non-abelian.
We show that we can use tachyon condensation to determine the position or the shape of D0-branes uniquely as a commutative region in spacetime together with non-trivial gauge flux on it,
even if the scalar fields are non-abelian. 
We use the idea of the so-called coherent state method
developed in the field of matrix models
in the context of the tachyon condensation.
We investigate configurations of noncommutative D2-brane made out of D0-branes as examples. In particular, we examine a Moyal plane and a fuzzy sphere in detail,
and show that whose shapes are commutative $\real^2$ and $S^2$, respectively, equipped with uniform magnetic flux on them. 
We study the physical meaning of this commutative geometry made out of matrices,
and propose an interpretation in terms of K-homology.\\

\end{abstract}


\end{titlepage}

\renewcommand{\thefootnote}{\arabic{footnote}}
\setcounter{footnote}{0}
\addtocounter{page}{1}

\tableofcontents

\section{Introduction}

D-branes in superstring theory are dynamical hypersurfaces in spacetime
on which gauge fields and transverse scalar fields live. 
On a single D-brane, the transverse scalar fields represent the displacement of the worldvolume in spacetime.
However, this interpretation cannot be applied naively for a stack of $N$ D-branes, 
since the scalar fields take values in $N\times N$ hermitian matrices,
which are not mutually diagonalizable in general.
Soon after the discovery of D-branes, 
the idea that such non-commuting scalar fields represent noncommutative (NC) geometry \cite{Witten1996,Connes1997} came out.
It is most readily seen 
by the matrix quantum mechanics for multiple D0-branes \cite{Banks1996}
or the matrix model \cite{Ishibashi1996}, 
which is (at least formally) seen as a model for D-instantons.
According to this matrix geometry picture, 
various NC configurations of scalar fields, representing 
NC spaces such as the Moyal plane {\cite{Aoki2000}} and the fuzzy sphere \cite{Myers1999} are considered.
In these examples, the non-abelian scalar fields on 
lower dimensional D-branes make the system couple to the RR 3-form
potential due to the Myers term
and the effective theories on them become NC gauge theories. 
Here there appears a puzzle: 
Such a NC worldvolume lives in usual commutative spacetime
while a NC space cannot be embedded into commutative spacetime 
in a usual sense in general. 
Therefore, the position or the shape of a NC D-brane 
in the commutative spacetime is far from obvious in particular. 

This problem has been discussed from various {viewpoints}. 
In {\cite{Hashimoto2004}}, the position of NC D-brane systems 
is estimated as the distribution of D-brane charges by using 
the D-brane charge density formula given in \cite{Taylor1999,Taylor1999a}.
The original charge density formula is improved 
by assuming that fuzzy sphere configurations
have single spherical shell structures, 
which gives a consistent improvement of the formula. 
This suggests that the worldvolume of NC D-brane system has a definite 
shape in the spacetime.

The way to determine the shape of D-brane system is not unique. 
The original interpretation that the diagonal elements of the scalar fields 
express the position of the worldvolume in spacetime 
has been generalized in \cite{Azeyanagi2009},
where the authors discuss that the position of the worldvolume should be 
identified by taking the ``almost diagonal gauge''  \cite{Hotta1998}
of the scalar fields. 

In \cite{Berenstein2012}, 
another interesting method of defining the shape of NC 
D-branes was proposed. In this method, in addition to NC D-branes, 
one introduces a probe D0-brane and considers open strings 
connecting the probe brane and the NC D-branes. 
The point is that the lowest energy of an open string is 
always proportional to the length of the string. 
Then, moving the position of the probe brane, 
one can find massless modes of open string only when the probe brane hits the 
NC branes so that the length of one of the open strings becomes zero.
Thus, the set of all possible positions of the probe brane such that
the open strings have massless modes can be interpreted as 
the shape of the NC branes. 
The energy of the open string can be measured by using 
a Dirac operator on the open strings and thus the shape of NC branes 
is defined as loci of zeros of the Dirac operator.
See also \cite{DeBadyn2015,Karczmarek2015} for analysis of this method.

The relation between NC and commutative geometries has been further developed 
as a mathematical correspondence between commutative geometry and matrices. 
In {\cite{Ishiki2015}}, a systematic way to extract a commutative space 
from a given configuration of matrices has been developed. 
In this approach, a Hamiltonian operator plays an important role, 
which is assumed to include matrices accompanied 
by coordinates of a Euclidean space $\real^n$ as parameters. 
The commutative manifold living in $\real^n$ is identified as 
loci of zero eigenstates of the Hamiltonian
and some geometrical quantities such as Poisson structures and Riemannian metrics 
can also be extracted by the coherent {states} \cite{Ishiki2015}
(See also \cite{Ishiki2016}). 
Although the large $N$ limit of the matrices has been considered in {\cite{Ishiki2015}}, 
it has been pointed out in \cite{Schneiderbauer2016} that this idea works even 
at finite $N$ with the use of the quasi coherent states, 
and it has been discussed that a Dirac-like operator 
can play the same role as the Hamiltonian.  
Interestingly enough, the obtained formulation is deeply related 
to that developed in the context of 
the superstring theory discussed in \cite{Berenstein2012}.  
We thus refer to the method developed in 
\cite{Berenstein2012,Ishiki2015,Schneiderbauer2016}
collectively as ``the coherent state method" hereafter.

In this paper, we point out that the coherent state method also
plays important roles in the context of 
the tachyon condensation in the superstring theory {\cite{Sen1998}}.
The basic idea is to identify the Dirac-like operator in the coherent state 
method with a tachyon profile on a system of unstable D-branes.
{With this identification, the coherent state method can be interpreted as the tachyon condensation,}
and the resultant commutative manifolds can be regarded as 
D-branes living in the commutative spacetime. 
The advantage of this interpretation is two-fold: 
First, the parameter space $\real^n$ in the coherent state method 
can be interpreted as a worldvolume of this unstable D-branes. 
Second, it gives a clear reason why the ground state should be chosen 
to extract the commutative worldvolume. 

Technically our analysis in this paper is an application 
of the technique developed so far \cite{Terashima2005}.
This method has been applied to realize the Nahm construction of monopoles and 
the ADHM construction of instantons \cite{Hashimoto2005,Hashimoto2006} and/or 
to realize a spherical D-brane \cite{Asakawa2017}. 
In the latter case, 
a system of $2$ non-BPS D3-branes is considered, where 
a tachyon profile $T$ representing a D0-brane is deformed by a constant shift.
By diagonalizing the tachyon $T$, 
the system is shown to condensate to a spherical D2-brane with 
a gauge flux of the unit monopole charge. 
Since the diagonalization is just the change of basis, 
the original deformed D0-brane and the spherical D2-brane 
with flux are unitary equivalent.
This construction is similar but different from the well-known Myers dielectric D2-brane. 
In the former case, a D2-brane is made out of a single D0-brane 
and its worldvolume is a commutative $S^2$,
while in the latter case a D2-brane is made out of multiple D0-branes 
and its worldvolume is a fuzzy sphere. 
In this paper, we apply the tachyon condensation to the latter case 
and show that the fuzzy sphere 
has an equivalent expression to a system on a commutative sphere.
For the latest result of the related topic, 
see \cite{Terashima}, 
which has some overlap with the present paper and 
has been appeared on arXiv at the same time with the present paper. 

The organization of this paper is as follows. 
In the section 2, 
{we consider the system of $k$ D0-branes with matrix-valued scalar fields on them, in terms of tachyon field of $2k$ non-BPS D3-branes.}
By using the idea of the coherent state method {in this setting,} we claim that the 
{shape of D0-branes is a commutative region $M$ of spacetime, and} is determined uniquely by the zeros of the tachyon field.
We {also} explain a general mechanism of appearing a gauge flux {on $M$}.
In the section 3, we apply the method to 
{NC D2-branes on} the Moyal plane and the fuzzy sphere,
{that are made of D0-branes with the Myers term.}
We identify the shapes of these systems as {\it commutative} $\real^2$ and $S^2$, respectively.
In the section 4, we discuss
{the meaning of the shapes more closely and propose an interpretation
in terms of K-homology.}
The section 5 is devoted to conclusion and discussion.


\section{Geometry from Matrices by Tachyon Condensation}

\subsection{Multiple D0-branes in non-BPS D-branes}

Consider a system of $N$ non-BPS D3-branes whose worldvolume is $\real\times \real^3$ in $10$-dimensional 
Minkowski spacetime.
The effective action is a $U(N)$ gauge theory 
coupled with $6$ transverse scalar fields and a tachyon field.
In this paper, we focus on static configurations of the tachyon field only.
{We also restrict the gauge connection to be trivial.}
In this setting, 
D3-branes are rigid 
and its spatial worldvolume is identified with a part of spacetime.
We set the spatial coordinates $\boldsymbol{x}=(x^1,x^2,x^3)$.
Moreover, the Chan-Paton bundle, 
a complex vector bundle over $\real^3$ with the fiber $\complex^{N}$, 
is trivial $\real^3 \times \complex^{N}$ because of no gauge field.
The tachyon field $T(\boldsymbol{x})$ is a hermitian $N\times N$-matrix valued scalar field.

Our argument below does not rely on the explicit form of the action, but for definiteness,
we assume the tachyon potential has the form $V(T)=e^{-T^2}$ 
 (i.e., we assume BSFT-type theory \cite{Takayanagi2000,Kraus2001}).
Because it is unstable around the (false) vacuum $T=0$, the tachyon condensation occurs.
At the true vacuum $T=u 1_N$ ($u\to \infty$), non-BPS D3-branes disappear.
In addition, lower dimensional D-branes can be realized as solitonic configurations
\cite{Sen1998}. 

Among them, let us consider $k$ D0-brane configuration with fluctuations.
We take $N=2k$ and set the tachyon profile as  
\bea
T(\boldsymbol{x})=u \boldsymbol{\sigma} \cdot (\boldsymbol{x}-\Phi)=u\mat{x^3-\Phi^3,\bar{z}-\bar{\Phi},z-\Phi,-x^3+\Phi^3},
\label{generic tachyon}
\ena
where
$\boldsymbol{\sigma}=(\sigma^1,\sigma^2,\sigma^3)$ is a set of Pauli matrices
and $\Phi=(\Phi^1,\Phi^2,\Phi^3)$ is a collection of transverse scalar fields on $k$ D0-branes,
that are $k\times k$ hermitian matrices.
In the second expression, we used complex notation with 
$z=x^1+ix^2$ and $\Phi=\Phi^1+i\Phi^2$.
Note that $x^i$ should be understood as $x^i \otimes \mathbf{1}_k$ more precisely.

The tachyon profile \eqref{generic tachyon} without fluctuation, $\Phi=0$, 
represents indeed $k$ D0-branes sitting at the origin $\boldsymbol{x}=0$ in the limit $u\to \infty$, 
which is known as the ABS construction {\cite{Atiyah1964, Witten1998}}.
This is essentially seen by the tachyon potential
\bea
V(T)= e^{-T^2}=e^{-u^2 |\boldsymbol{x}|^2} \otimes  \mathbf{1}_{2k},
\ena 
which is proportional to the delta function $\delta(\boldsymbol{x})$ in the $u\to \infty$ limit.
Thus, under the tachyon condensation, the spatial worldvolume $\real^3$ reduces to 
the origin, leaving the point-like defect.
This fact is most rigorously shown by using boundary states 
(see for example {\cite{Asakawa2003}}):
The boundary state for non-BPS D3-branes with this tachyon profile added on as a boundary 
interaction reduces to the boundary state for $k$ D0-branes in the limit $u\to \infty$, 
with the correct tension and the RR-charge.
Even adding fluctuations, the profile \eqref{generic tachyon} 
reduces to $k$ D0-branes with transverse scalars,
where scalar fields appear as a boundary interaction.
The resulting effective action $S_{\rm D0}[\Phi]$ for $k$ D0-branes is given by the DBI action and the Chern-Simons term, 
{which in particular includes the Myers term \cite{Myers1999}.}
In the opposite way, a matrix model $S_{\rm D0}[\Phi]$ can be embedded into the theory of non-BPS D3-branes.
This explains why the tachyon field appears in considering 
the shape of D-branes with non-commuting scalar fields.

Note that, in this treatment, the condensation itself is obtained without matrix scalar fields $\Phi$,
and $\Phi$ are turned on afterwards as perturbation.
Equivalent but more direct way is to consider the condensation of the profile 
\eqref{generic tachyon} with $\Phi$.
The resulting defect should be the deformation of the point-like defect by matrices $\Phi$.
Indeed, as shown in \cite{Asakawa2017}, a deformation of the single 
D0-brane ($k=1$) profile drastically changes the condensation defect 
to a spherical D2-brane%
\footnote{It is a deformation of \eqref{generic tachyon} with $\Phi=0$ by a constant shift 
and thus different from $\Phi$ here.}.
Our claim in this paper is that the position or the shape of 
{D-branes} is 
determined by diagonalizing the tachyon field $T${,} 
not the scalar fields $\Phi$ themselves. 

In the following, we will consider such configurations of matrix scalar fields $\Phi$ that represent noncommutative D2-branes as typical examples. In particular, we investigate a NC plane and a fuzzy sphere in detail.
By embedding D0-branes into non-BPS D3-branes,
we will see that the spatial worldvolume $\real^3$ shrinks 
to a commutative $2$-dimensional space after the tachyon condensation.
Moreover, this process induces a non-trivial gauge fields inevitably, 
whose field strength carries the D0-brane charge $k$.

\subsection{Tachyon condensation and gauge flux production}

Before treating explicit examples, we describe the schematic structure 
of the tachyon condensation for the configuration \eqref{generic tachyon}.
Technically, the analysis is the same as the coherent state method mentioned in the introduction.
We also explain how a non-trivial $U(1)$ gauge flux is induced from the tachyon condensation. 
{In order to consider the case of not only finite but also infinite $N$, we}
formulate the problem in terms 
of Hilbert spaces and projective modules.

\paragraph{Tachyon condensation}

The Chan-Paton bundle for $N$ non-BPS D3-brane in our setting 
is a trivial complex vector bundle $E=\real^3 \times \complex^N$ over $\real^3$,
whose typical fiber $\complex^N$ is a Hilbert space.
Then, the space of sections of the Chan-Paton bundle is a free module ${\cal A}^N$ of rank $N$,
with ${\cal A}=C^\infty (\real^3)$.
Denote an orthonormal basis (ONB) for $\complex^N$ as $\Ket{a}$ ($a=0,1,2,\ldots,N-1$).
Then, a generic section is written as
\bea
\Ket{\psi(\boldsymbol{x})}=\sum_{a=0}^{N-1} \psi^a (\boldsymbol{x}) \Ket{a},
\quad \psi^a (\boldsymbol{x})=\Bracket{a}{\psi(\boldsymbol{x})}\in {\cal A}.
\ena

The tachyon field $T(\boldsymbol{x})$ is an operator-valued function on $\real^3$.
It is an element of the endomorphism ${\rm End}(E)$ and 
is written as 
$T(\boldsymbol{x})=\sum_{a,b}\Ket{a}T^a_b(\boldsymbol{x})\Bra{b}$.
According to \eqref{generic tachyon}, 
we assume that each matrix element $T^a_b$ is at order $u$,
and the limit $u\to \infty$ will be taken.
Note that in this profile \eqref{generic tachyon}, the matrices $\Phi$ act at each 
$\boldsymbol{x}$ (not only at the origin).

In order to extract the condensation defect, 
we need to diagonalize the potential $V(T)=e^{-T^2}$, 
or the tachyon field $T(\boldsymbol{x})$ itself at each point $\boldsymbol{x}$ on $\real^3$.
Any hermitian matrix can be diagonalized by a unitary matrix.
In an infinite dimensional Hilbert space and operators acting on it,
the corresponding notion is the spectral decomposition.
Assume the spectral decomposition at each point $\boldsymbol{x}$,
\bea
 T(\boldsymbol{x})=U(\boldsymbol{x})T_0(\boldsymbol{x})U(\boldsymbol{x})^\dagger, 
 \quad  T_0(\boldsymbol{x})=\sum_a \Ket{a}t_a (\boldsymbol{x})\Bra{a},
\label{spectral decomposition of T}
\ena
we find the eigenstates for the tachyon field as 
\bea
T(\boldsymbol{x})\Ket{\psi_a(\boldsymbol{x})}
=t_a (\boldsymbol{x})\Ket{\psi_a(\boldsymbol{x})}, 
\quad \Ket{\psi_a(\boldsymbol{x})}=U(\boldsymbol{x})\Ket{a}.
\ena
Here an eigenstate $\Ket{\psi_a (\boldsymbol{x})}$, which is a section 
of the Chan-Paton bundle, and the unitary operator $U (\boldsymbol{x})$ are position dependent.
In more familiar term, \eqref{spectral decomposition of T} 
is a gauge transformation of the tachyon field.
In general, eigen-functions $t_a(\boldsymbol{x})$ are $u$-dependent but 
$U (\boldsymbol{x})$ is $u$-independent (see the examples below).

Then, the tachyon potential is written as 
\bea
e^{-T^2}
=U(\boldsymbol{x})e^{-T_0(\boldsymbol{x})^2}U(\boldsymbol{x})^\dagger 
=\sum_{a} U(\boldsymbol{x})\Ket{a} e^{-t_a (\boldsymbol{x})^2}\Bra{a}U(\boldsymbol{x})^\dagger.
\ena
This shows that, at each $\boldsymbol{x}$, the component with $t_a(\boldsymbol{x})\ne 0$ tends to $0$ in the limit $u\to \infty$.
That is, the tachyon potential picks up the tachyon zero modes at each point.
For example%
\footnote{This is just a working assumption. 
More general situations are discussed in \S\ref{sec:comments}.}, 
if only one component is the zero mode $t_0(\boldsymbol{x})=0$ 
for any $\boldsymbol{x}$, then 
\bea
e^{-T^2} \to P(\boldsymbol{x})
=U(\boldsymbol{x})\Ket{0}\Bra{0}U(\boldsymbol{x})^\dagger.
\label{potential to P}
\ena
In this case, all the excited state $\Ket{a}$ ($a\ne 0$) are annihilated under the tachyon condensation.
Note that $P_0=\Ket{0}\Bra{0}$ is a rank $1$ projection operator acting on the typical fiber,
and the $P(\boldsymbol{x})$ is unitary equivalent to $P_0$.
It means that 
the tachyon condensation picks up a $1$-dimensional 
subspace $U(\boldsymbol{x})\Ket{0}$ from the $N$-dimensional fiber
at each point $\boldsymbol{x}$. 
More generally, it may happen that $t_0 (\boldsymbol{x})=0$ for some region 
$M\subset \real^3$, 
but $t_0 (\boldsymbol{x})\ne0$ otherwise.
In this case, the tachyon potential also projects out the region $M$.
Schematically, 
\bea
e^{-T^2} \to \delta(M) P(\boldsymbol{x})
=\delta(M)U(\boldsymbol{x})\Ket{0}\Bra{0}U(\boldsymbol{x})^\dagger,
\label{potential to delta}
\ena
where $\delta(M) $ denotes a delta function distribution with its support on $M$.
The original information of a choice of matrices $\Phi$ in \eqref{generic tachyon} 
is transferred to 
two kinds of information, $\delta(M)$ and $P(\boldsymbol{x})$.

Note that this procedure is completely point-wise, 
and in general the unitary operator $U(\boldsymbol{x})$ is not globally defined 
as a smooth function on the whole $\real^3$.
For such cases, we may apply the procedure by considering patch-wise.
That is, choose an open covering $\{ {\cal U}_I\}$ of $\real^3$, 
such that the corresponding set of unitary operators $\{U_I (\boldsymbol{x})\}$ 
are defined smoothly on each ${\cal U}_I$.
For a point in the overlap ${\cal U}_I \cap {\cal U}_J$, there are two 
diagonalizations
but they give the same defect because the eigen-function $t_0(\boldsymbol{x})$ is 
gauge independent.
Then, the region $M$ is also given patch-wise by the union $M=\cup_I M_I$.
Our examples below are of the type \eqref{potential to delta}
with $M=\real^2$ and $M=S^2$.
In the latter case, the patch-wise condensation is needed.
In these cases, a defect after the condensation is interpreted as a D2-brane on $M$.  
What kind of $M$ appears depends of course on the choice of the matrices $\Phi$.

In summary, the tachyon condensation just picks up the zeros of the eigenfunction 
and as a result a defect remains on a region $M$.
This is technically the same as the coherent state method mentioned in the introduction.
In fact, the tachyon profile $T$ \eqref{generic tachyon} is exactly the
same as the {Dirac-like} operator
in the literature {\cite{Berenstein2012,DeBadyn2015,Karczmarek2015,Schneiderbauer2016}
and $T^2$ corresponds to the Hamiltonian in \cite{Ishiki2015,Ishiki2016}.}
Our claim in this paper is that the tachyon condensation gives a new
physical interpretation of this prescription, based on the dynamics of the 
non-BPS D-branes. Although we are working with static configurations, the condensation is essentially a dynamical process and the zero modes survive as a result of the dynamics. 
This is in contrast with the previously proposed interpretations in \cite{Berenstein2012,Ishiki2015,Schneiderbauer2016}
of the coherent state method, which are based on statics.

\paragraph{Gauge flux production}

Tachyon potential of the form \eqref{potential to P} or \eqref{potential to delta} 
induces a $U(1)$-flux.
We here briefly describe the {mechanism} of this {effect}.
For more detail we refer the reader to \cite{Asakawa2017}.

On the Chan-Paton bundle, the tachyon potential \eqref{potential to P}
plays the role of a projection operator $P(\boldsymbol{x})$, 
which picks up a subspace $U(\boldsymbol{x})\Ket{0}$ at each fiber. 
This defines a projective module $P{\cal A}^N$, which is identified as 
the space of sections of a line bundle on $\real^3$.
Since $U(\boldsymbol{x})$ is a unitary operator, 
$U(\boldsymbol{x})\Ket{a}$ forms an orthonormal basis at each fiber.
We may then write a generic element of the free module ${\cal A}^N$ in this new basis as
\bea
\Ket{\psi(\boldsymbol{x})}=\sum_a \psi^a (\boldsymbol{x}) U(\boldsymbol{x})\Ket{a}.
\ena
An element of the projective module $P{\cal A}^N$ is then given by
\bea
P(\boldsymbol{x})\Ket{\psi(\boldsymbol{x})}
=\psi^0 (\boldsymbol{x}) U(\boldsymbol{x}) \Ket{0}.
\ena
Since $P(\boldsymbol{x})$ depends on $\boldsymbol{x}$, 
the exterior derivative $d$ does not preserve the module $P{\cal A}^N$ in general.
This leads to the notion of connections.
A natural connection on $P{\cal A}^N$, called the Grassmannian connection, is defined by $\nabla=P\circ d$, which acts as
\bea
Pd(P\Ket{\psi})
&=Pd (\psi^0 U\Ket{0}) \nn
&=P(d\psi^0 U\Ket{0}+\psi^0 dU \Ket{0}) \nn
&=d\psi^0 U\Ket{0} +\psi^0 U\Ket{0}\Bra{0}U^\dagger dU \Ket{0} \nn
&=\left(d\psi^0 +iA \psi^0 \right) U\Ket{0},
\ena
where 
\bea
iA(\boldsymbol{x})=\Bra{0}U(\boldsymbol{x})^\dagger dU(\boldsymbol{x})\Ket{0}.
\label{def of gauge field}
\ena
In components, we obtain $\psi^0 \to d\psi^0 +iA \psi^0$ the covariant exterior derivative
on the line bundle with a $U(1)$ gauge potential $A$.
In the case that the tachyon potential has the form \eqref{potential to delta}, 
this gauge field is also confined to the region $M \subset \real^3$ because of the 
delta-function distribution. 
In this case, \eqref{def of gauge field} has components only along $M$ 
 (for the proof, see \cite{Asakawa2017}).
If $U(\boldsymbol{x})$ is not globally defined 
and there are two different diagonalizations $\{U_I(\boldsymbol{x})\}$ at a point,  
then two gauge potentials of the form \eqref{def of gauge field} are related by 
$U(1)$ transition function.

This gauge potential should possess a non-trivial $U(1)$-flux
\bea
iF(\boldsymbol{x})=idA(\boldsymbol{x})
= \Bra{0}dU(\boldsymbol{x})^\dagger dU(\boldsymbol{x})\Ket{0},
\label{def of field strength}
\ena
in our setting. {Note that} the D0-brane charge $k$ for 
the original $k$ D0-brane system described by $\Phi$ should be maintained
as a magnetic flux of charge $k$ in the D2-brane on $M$.
We will see this explicitly in the next section. 

Note that the induced gauge potential $A$ in \eqref{def of gauge field} 
can also be seen as a Berry connection,
if we regard the base space $\real^3$ as a parameter space of the single Chan-Paton space $\complex^N$.
This viewpoint is appeared in \cite{Hashimoto2005,Hashimoto2006} in the context of the tachyon condensation, and in \cite{Ishiki2015} in the coherent state method.

\section{Examples for NC D2-branes}

The $k$ D0-brane solution with fluctuation is given by \eqref{generic tachyon}
\bea
T(\boldsymbol{x})=u \boldsymbol{\sigma} \cdot (\boldsymbol{x}-\Phi)=u\mat{x^3-\Phi^3,\bar{z}-\bar{\Phi},z-\Phi,-x^3+\Phi^3},
\label{generic tachyon 2}
\ena
where $z=x^1+ix^2$ and $\Phi=\Phi^1+i\Phi^2$.
Note that $x^i$ should be understood as $x^i \otimes \mathbf{1}_k$ more precisely.
$\Phi^i$ ($i=1,2,3$) are transverse scalar fields on $k$ D0-branes and are $k\times k$ hermitian matrices.
In this section, we consider examples that matrices $\Phi^i$ represent NC D2-branes, 
on a Moyal plane and a fuzzy sphere.
The shape of these branes are a commutative $\real^2$ and $S^2$, respectively.

\subsection{Moyal plane}

A NC D2-brane on the Moyal plane can be made out of $k$ D0-branes, 
if the scalar field has the profile
\bea
\Phi^1=\hx^1, \quad \Phi^2=\hx^2, \quad \Phi^3=0,
\label{Moyal sol}
\ena
where $\hx^1$ and $\hx^2$ are coordinates on a Moyal plane 
satisfying $[\hx^1,\hx^2]=i\theta$.
By defining the creation/annihilation operators by
\bea
\ha=\frac{1}{\sqrt{2\theta}}(\hx^1 +i\hx^2), \quad 
\had=\frac{1}{\sqrt{2\theta}}(\hx^1 -i\hx^2),
\ena
the scalar fields \eqref{Moyal sol} are rewritten in complex notation as
\bea
\Phi=\Phi^1+i\Phi^2=\sqrt{2\theta}\ha, \quad \Phi^3=0.
\ena
In order to realize them, it is necessary to take $k\to \infty$ 
and replace matrices $\Phi^i$ with operators acting on the Hilbert space $\ell^2 (\natural)$.

By inserting \eqref{Moyal sol} into \eqref{generic tachyon 2}, the tachyon profile becomes
\bea
T(\boldsymbol{x})
=u \boldsymbol{\sigma} \cdot (\boldsymbol{x}-\Phi)
=u\mat{x^3,\bar{z}-\sqrt{2\theta}\had,z-\sqrt{2\theta}\ha,-x^3}.
\label{T for Moyal}
\ena
It acts on the Chan-Paton bundle with typical fiber to be the Hilbert space 
${\cal H} =\ell^2 (\natural) \otimes \complex^2$.
Let $\{\Ket{n,\epsilon}|n=0,1,2,\cdots, \epsilon =\pm\}$ be its ONB,
where {$n$ and} $\epsilon$ denote
{the eigenstate for the number operator $\hat{N}=\had\ha$ and the}
eigenvalues {of the} Pauli matrix $\sigma_3${, respectively.}
That is, two component vectors
\bea
\Ket{n,+}=\vect{\Ket{n},0}, \quad \Ket{n,-}=\vect{0,\Ket{n}},
\ena
give the basis of $\cal{H}$, where $\Ket{n}$ is the ONB of $\ell^2(\natural)$.

\paragraph{Condensation}
We will now study the tachyon condensation of this profile \eqref{T for Moyal}.
To this end, we use the displacement operator for $\alpha \in \complex$, 
\bea
D(\alpha)=e^{\alpha \had-\bar{\alpha}\ha}
=e^{-|\alpha|^2 /2} e^{\alpha \had} e^{-\bar{\alpha}\ha},
\label{displacement op}
\ena
which is a unitary operator 
and defines a coherent state $\Ket{\alpha}=D(\alpha)\Ket{0}$ {\cite{Gazeau}}.
The basic properties are
\bea
D(\alpha)\ha D(\alpha)^\dagger =\ha -\alpha, 
~~D(\alpha)\had D(\alpha)^\dagger =\had -\bar{\alpha}.
\label{displacing}
\ena
By using these properties, $z$-dependence in \eqref{T for Moyal} is extracted as
\bea
T(\boldsymbol{x})=u U(z)\mat{x^3,-\sqrt{2\theta}\had,-\sqrt{2\theta}\ha,-x^3}U^\dagger(z),
\label{displaced T}
\ena
where the unitary operator $U(z)$ is given by
\bea
U(z) &=\mat{D(\alpha),0,0,D(\alpha)}, \quad
\alpha=\frac{z}{\sqrt{2\theta}}.
\label{def of alpha}
\ena
Under the tachyon condensation $u\to\infty$, 
the surviving mode under the condensation is zero eigenstates of $T^2(\boldsymbol{x})$,
\bea
T^2(\boldsymbol{x})=u^2U(z)
\mat{(x^3)^2 + 2\theta \hat{N}, 0,0,(x^3)^2 + 2\theta(\hat{N}+1)}U(z)^\dagger.
\ena
It exists only for $x^3=0$.
Since $\hat{N}$ has the spectrum $\{n=0,1,2,\ldots\}$, 
$T^2(\boldsymbol{x})$ has a zero mode of the form
\bea
U(z)\Ket{0,+}=\vect{D(\alpha)\Ket{0},0},
\label{Moyal zero mode}
\ena
at each point with arbitrary $z=x^1+ix^2 $ and $x^3=0$.
The tachyon potential reduces to the projection operator onto this zero mode:
\bea
&e^{-T^2} \xrightarrow{u\to\infty}
\frac{u}{\sqrt{\pi}}\delta(x^3) P(z), \quad
P(z)=U(z)\mat{\Ket{0}\Bra{0},0,0,0}U(z)^\dagger.
\label{NC R2 projection}
\ena
Here we used the fact $e^{-u^2 (x^3)^2} \to \frac{u}{\sqrt{\pi}}\delta(x^3)$ in the limit
$u\to \infty$. 

From the delta function, we see that the remnant of this condensation 
is the real $2$-dimensional surface $M=\real^2=\complex$,
which is considered as a spatial worldvolume of a D2-brane.
We emphasize that the {obtained} worldvolume 
{parameterized by $z$ and $\bar{z}$ is} commutative, 
although we start with a Moyal plane configuration.
On the other hand, the projection operator $P(z)$ of the Chan-Paton bundle
picks up a coherent state $D(\alpha)\Ket{0}$ at each point $z$ on $M$.
Because it is $1$-dimensional subspace at each fiber, the Chan-Paton bundle
reduces to a line bundle on $M$.
It means a single D2-brane with the gauge group $U(1)$.
Moreover, because the fiber $D(\alpha)\Ket{0}$ smoothly depends on the base space 
(recall \eqref{def of alpha}), this line bundle is non-trivial.
This information is encoded in the unitary operator $U(z)$,
and we see a further consequence on the gauge flux in the following. 
  
\begin{figure}[h]
\begin{center}
	\includegraphics[scale=0.7]{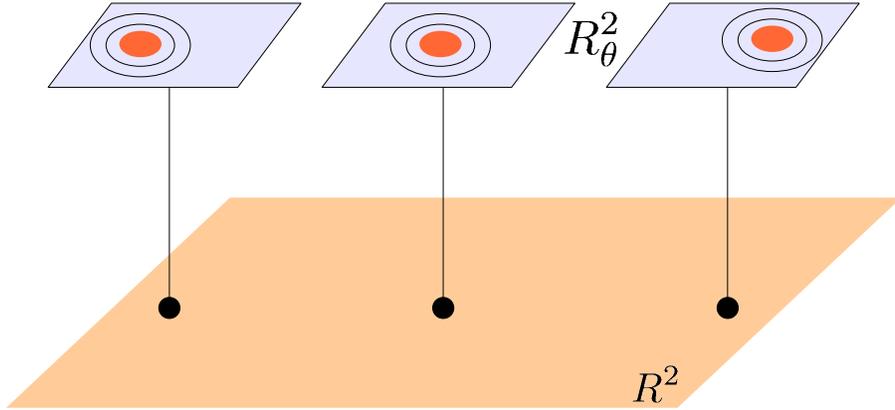}
\caption{\small 
The large plane represents 
the base space $M=\real^2$. After the tachyon condensation, 
at each point on the base space, 
we have $D(\alpha)\Ket{0}$ as the fiber of the line bundle on $M$.
The wave packets of $D(\alpha)\Ket{0}$, which
have area $2\pi \theta$,
are schematically drawn as the colored blobs on the smaller 
planes.
} 
\label{fig:Moyal}
\end{center}
\end{figure}

It is worth emphasizing that this result is completely different from 
perturbative picture of multiple D0-branes,
where a D0-brane is sitting at the origin but 
{is fluctuating around the origin to the ``directions'' of} 
the non-commuting scalar fields $\Phi$, that is, a single Moyal plane.
In our picture, matrices $\Phi$ originally give a family of Moyal planes on $\real^3$
as a Chan-Paton bundle of the non-BPS D3-branes,
which however reduces to a line bundle on $M=\real^2$ by the tachyon condensation.
The schematic picture is given in Figure \ref{fig:Moyal}.

\paragraph{Eigenstates}
For completeness, we here diagonalize \eqref{T for Moyal}.
We give the solution for the eigenvalue problem
\bea
T(\boldsymbol{x})\Ket{\psi_{n,\epsilon}(\boldsymbol{x})}
=t_{n,\epsilon}(\boldsymbol{x}) \Ket{\psi_{n,\epsilon}(\boldsymbol{x})}.
\ena
Note that states of the form $D(\alpha)\Ket{n,\epsilon}$ give an ONB, 
since the displacement operator is a unitary operator.
Acting $T$ on these states, 
it is easy to recognize that $D(\alpha)\Ket{0,+}$ is already an eigenstate 
$TD(\alpha)\Ket{0,+}=u x^3 D(\alpha)\Ket{0,+}$.
Thus we write it as $\Ket{\psi_{0,+}}=D(\alpha)\Ket{0,+}$ with 
$t_{0,+}(\boldsymbol{x})=u x^3$.
Next, for a fixed $n$ ($n \ge 1$), two states 
$D(\alpha)\Ket{n,+}$ and $D(\alpha)\Ket{n-1,+}$ 
form a doublet under $T$, since 
\bea
&T(\boldsymbol{x})D(\alpha)\Ket{n,+}
=uD(\alpha)\left(x^3 \Ket{n,+}-\sqrt{2\theta n}\Ket{n-1,-}\right),\nn
&T(\boldsymbol{x})D(\alpha)\Ket{n-1,-}
=uD(\alpha)\left(-x^3 \Ket{n-1,-}-\sqrt{2\theta n}\Ket{n,+}\right).
\ena
On this doublet, $T$ is effectively a matrix for each $n$,
\bea
T^{(n)}=\mat{x^3,-\sqrt{2\theta n},-\sqrt{2\theta n},-x^3}
=x^3 \sigma_3 -\sqrt{2\theta n} \sigma_1,
\ena
{which is} easily diagonalized by the unitary matrix 
\bea
W^{(n)} =\frac{1}{\sqrt{2|T^{(n)}|(|T^{(n)}|+x^3)}}
\mat{|T^{(n)}|+x^3,\sqrt{2\theta n},-\sqrt{2\theta n},|T^{(n)}|+x^3},
\ena
where $|T^{(n)}|=\sqrt{(x^3)^2+2\theta n}$.
The eigenvalues are $t_{n,\epsilon}(\boldsymbol{x})=u\epsilon |T^{(n)}|$
and the {corresponding} eigenstates are 
\bea
\Ket{\psi_{n,+}}=W^{(n)}_{++}D(\alpha)\Ket{n,+}+W^{(n)}_{+-}D(\alpha)\Ket{n-1,-},\nn
\Ket{\psi_{n,-}}=W^{(n)}_{-+}D(\alpha)\Ket{n,+}+W^{(n)}_{--}D(\alpha)\Ket{n-1,-}.
\ena
We {can express all the eigenstates} 
as $\Ket{\psi_{n,\epsilon}}=WD(\alpha)\Ket{n,\epsilon}$,
by defining $W$ as $1$ on $\Ket{0,+}$ and $W^{(n)}$ on the doublet at $n$ as above.
These states are orthonormal 
$\Bracket{\psi_{n,\epsilon}}{\psi_{n',\epsilon'}}=\delta_{nn'}\delta_{\epsilon\epsilon'}$,
since $W^\dagger W=1$.

In summary, the set of eigenstates consists of a ground state (singlet) 
$\Ket{\psi_{0,+}}=D(\alpha)\Ket{0,+}$ with its eigenvalue 
$t_{0,+}(\boldsymbol{x})=u x^3$,
and the family of doublets $\Ket{\psi_{n,\epsilon}}$ ($n\ge 1)$ with eigenvalues 
$t_{n,\epsilon}(\boldsymbol{x})=u\epsilon\sqrt{(x^3)^2+2\theta n}$.
Under the tachyon condensation, all the doublets are annihilated, 
because $t_{n,\epsilon}(\boldsymbol{x})\ne0 $ for all $\boldsymbol{x}$,
while the singlet survives on the plane $x^3=0$ as states in \eqref{Moyal zero mode}.
Note that the mixing between states $\Ket{n}$ and $\Ket{\pm}$ is inevitable.
This structure cannot be seen by the part $\Ket{n}$ only 
(i.e, Chan-Paton space for D0-branes).

\paragraph{Gauge flux}
The tachyon potential \eqref{NC R2 projection} defines the projective module
or equivalently a complex line bundle over $M=\real^2$.
The corresponding $U(1)$ gauge connection is given by the 
Grassmannian connection
according to \eqref{def of gauge field}.
In the present case, the $U(1)$ gauge field on $M=\real^2$ is given by
\bea
iA(z,\bar{z})
&=\Bra{0,+}U^\dagger (z)dU(z)\Ket{0,+}
=\Bra{0}D^\dagger (\alpha)dD(\alpha)\Ket{0}.
\ena
After some calculations, we find 
\bea
A=-\frac{i}{4\theta}(\bar{z}dz -z d\bar{z})
=\frac{1}{2\theta}(x^1dx^2-x^2dx^1).
\ena
{(See Appendix \ref{appendix1} for derivation.)}
The corresponding field strength on $M$ is given by
\bea
F=dA=\frac{i}{2\theta}dz \wedge d\bar{z}=\frac{1}{\theta}dx^1\wedge dx^2.
\label{gauge flux for Moyal}
\ena
A uniform magnetic flux on a D2-brane is interpreted as the D0-brane charge density,
and its presence indicates that the resulting system is a bound state of D2 and D0-branes,
where D0-branes are dissolved into a D2-brane.
In fact, in the Chern-Simons term for a D2-brane, 
the coupling to the RR $1$-form is 
\bea
\frac{1}{2\pi}\int_{\real^2}F
=\frac{{\rm Vol}(\real^2)}{2\pi \theta}.
\ena
It says that there is a dissolved D0-brane per unit volume $2\pi \theta$.
Therefore, the original information on a Moyal plane 
is converted to a commutative plane with a uniform magnetic flux.

This equivalence between commutative and noncommutative descriptions 
of the D2-D0 bound states is first shown in {\cite{Ishibashi1999}}, in terms of boundary states.
We here reproduce the same result within the effective theory on non-BPS D3-branes,
but the equivalence is realized in a more direct way.
That is, once the D2-D0 bound states is represented in the tachyon profile, 
the equivalence is realized by the unitary transformation that diagonalizes the 
tachyon profile.

\subsection{Fuzzy sphere}

A NC D2-brane on a fuzzy sphere can be made out of $k$ D0-branes, 
if the scalar field has the profile
\bea
\Phi^i =\rho L_i ,\qquad [L_i,L_j]=i{\varepsilon_{ij}}^k L_k, 
\label{fuzzy sol}
\ena
where $\rho$ is a real parameter and $L_i$ ($i=1,2,3$) are $su(2)$ generators 
in {the} spin-$\ell$ irreducible representation {\cite{Madore1992}}.
Thus it is possible for $k\ge 2$.
We denote corresponding $k=2\ell +1$ states as 
$\Ket{m}$ ($m=-\ell, -\ell+1,\ldots, \ell-1, \ell$).
Because 
$\Phi^2=\rho^2 \boldsymbol{L}^2=\rho^2 \ell(\ell +1) {\bf 1}_k 
=\rho^2 \frac{k^2 -1}{4} {\bf 1}_k$,
a naive guess of the radius of this fuzzy sphere is $\rho \sqrt{\frac{k^2 -1}{4}}$.
We will compare it with the radius of $S^2$ obtained from the tachyon condensation below.

By inserting \eqref{fuzzy sol} into \eqref{generic tachyon 2}, 
the tachyon profile becomes
\bea
T(\boldsymbol{x})=u \boldsymbol{\sigma} \cdot (\boldsymbol{x}-\rho \boldsymbol{L}),
\label{S2 tachyon}
\ena
and its square leads to
\bea
T^2(\boldsymbol{x})
&=u^2 \left(|\boldsymbol{x}|^2 +\rho^2\boldsymbol{L}^2 -2\rho(\boldsymbol{x}\cdot \boldsymbol{L}) -\rho^2(\boldsymbol{\sigma}\cdot \boldsymbol{L})\right). 
\ena
Here, the ONB of the Chan-Paton Hilbert space 
${\cal H}=\complex^k \otimes \complex^2$ is given by 
$\{\Ket{m,\epsilon}| m=-\ell,\ldots, \ell, \epsilon=\pm\}$.

\paragraph{Condensation}
Here we study the tachyon condensation by diagonalizing $T$ in \eqref{S2 tachyon}.
{To this end, we examine the two terms in}
\eqref{S2 tachyon} separately {in detail}.
\begin{enumerate}
\item[a)] 
The term $\boldsymbol{\sigma} \cdot \boldsymbol{x}$ in \eqref{S2 tachyon} 
is independent of the choice of $\Phi$, and it can be diagonalized only
patch-wise {\cite{Asakawa2017}.}

First at the origin $\boldsymbol{x}=0$ in $\real^3$, 
this term does not contribute to $T$ and is already diagonal.
We then divide $\real^3$ except $\boldsymbol{x}=0$ into two regions:
\bea
&{\cal U}_N=\{\boldsymbol{x} \in \real^3 ~|~ |\boldsymbol{x}|+x^3\ne0 \}
=\{(r,\theta,\varphi) \in \real^3 ~|~r\ne 0, \theta\ne \pi\},\nn
&{\cal U}_S=\{\boldsymbol{x} \in \real^3 ~|~ |\boldsymbol{x}|-x^3\ne0 \}
=\{(r,\theta,\varphi) \in \real^3 ~|~r \ne 0, \theta\ne 0\},
\ena
where $\boldsymbol{x}=(x^1,x^2,x^3)$ and in the second expression 
the standard polar coordinates are used. 
Thus, ${\cal U}_N$ is $\real^3$ except for the negative $x^3$-axis,
while ${\cal U}_S$ is $\real^3$ except for the positive $x^3$-axis.
In each region ${\cal U}_N$ and ${\cal U}_S$, 
$\boldsymbol{\sigma} \cdot \boldsymbol{x}$ is diagonalized as
\bea
R^\dagger_{N/S}(\Omega)(\boldsymbol{\sigma} \cdot \boldsymbol{x})R_{N/S}(\Omega)
=|\boldsymbol{x}| \sigma_3,
\label{rotation of sigma x}
\ena
by the corresponding unitary matrix-valued function on $\real^3$
\bea
&R_N(\Omega)
=\frac{1}{\sqrt{2|\boldsymbol{x}| (|\boldsymbol{x}|+x^3)}}
\mat{|\boldsymbol{x}|+x^3, -\bar{z},z, |\boldsymbol{x}|+x^3},\nn
&R_S(\Omega)
=\frac{1}{\sqrt{2|\boldsymbol{x}| (|\boldsymbol{x}|-x^3)}}
\mat{\bar{z},-|\boldsymbol{x}|+x^3, |\boldsymbol{x}|-x^3,z}.
\label{AM}
\ena
They depend only on the angular coordinates $\Omega=(\theta,\varphi)$
and are written in the polar coordinates as
\bea
&R_N(\Omega)
=\mat{\cos {\textstyle \frac{\theta}{2}}, -\sin {\textstyle \frac{\theta}{2}}e^{-i\varphi}, 
\sin {\textstyle \frac{\theta}{2}}e^{i\varphi}, \cos {\textstyle \frac{\theta}{2}}},\quad
R_S(\Omega)
=\mat{\cos {\textstyle \frac{\theta}{2}}e^{-i\varphi}, -\sin {\textstyle \frac{\theta}{2}}, 
\sin {\textstyle \frac{\theta}{2}}, \cos {\textstyle \frac{\theta}{2}}e^{i\varphi}}.
\label{AM2}
\ena
The expression $R_N$ in \eqref{AM2} is familiar in quantum mechanics 
with the diagonalization of a spin with respect to the direction 
$\hat{\boldsymbol{x}} =\frac{\boldsymbol{x}}{|\boldsymbol{x}|}$,
if $S_i=\frac{\sigma_i}{2}$ is considered as the spin-$\frac{1}{2}$ representation.
But note that the diagonalization by $R_N$ is ill-defined at the south pole 
$\theta=\pi$.%
\footnote{
It is obvious in \eqref{AM} if $|\boldsymbol{x}|+x^3=0$, and 
in \eqref{AM2} it is ill-defined because $\varphi$ is undefined at $\theta=\pi$.}
In order to cover all the direction, we need another open set ${\cal U}_S$.

\item[b)]
The term $\boldsymbol{\sigma} \cdot \boldsymbol{L}$ in \eqref{S2 tachyon}
or more properly, the term $\boldsymbol{S} \cdot \boldsymbol{L}$ 
is similar to the spin-orbit interaction in quantum mechanics.
Thus, under the total spin $\boldsymbol{J}=\boldsymbol{L}+\boldsymbol{S}$, 
the tensor product representation
$[\ell]\otimes [\frac{1}{2}]$ decomposes into two irreducible 
representations $[\ell + \frac{1}{2}]\oplus [\ell - \frac{1}{2}]$.
Since $\boldsymbol{J}^2=\boldsymbol{L}^2+\boldsymbol{S}^2+2(\boldsymbol{S}\cdot \boldsymbol{L})$, 
the operator $\boldsymbol{\sigma}\cdot \boldsymbol{L}$ has the eigenvalue $\ell$ in all states in $[\ell + \frac{1}{2}]$,
while the eigenvalue $-(\ell+1)$ in all states in $[\ell - \frac{1}{2}]$.
This shows that two kinds of states should be mixed, in order to diagonalize 
$\boldsymbol{\sigma} \cdot \boldsymbol{L}$.
The ONB ($2\ell+2$ states) of $[\ell + \frac{1}{2}]$ are given by 
eigenstates of $J_3$ as
\bea
\Ket{m+\textstyle{\frac{1}{2}}}_{\ell + \frac{1}{2}}
=\alpha_m\Ket{m,+}
+\beta_m\Ket{m+1,-},
\ena
where $m=-\ell-1,-\ell,\ldots,\ell$, and 
\bea
\alpha_m=\sqrt{\frac{\ell+m+1}{2\ell+1}},\qquad
\beta_m=\sqrt{\frac{\ell-m}{2\ell+1}}.
\label{alphabeta}
\ena
Note that two particular states 
\bea
\Ket{\ell+\textstyle{\frac{1}{2}}}_{\ell + \frac{1}{2}}=\Ket{\ell,+},
\quad \Ket{-\ell-\textstyle{\frac{1}{2}}}_{\ell + \frac{1}{2}}=\Ket{-\ell,-},
\ena
exist in this representation.
On the other hand, the ONB ($2\ell$ states) of $[\ell - \frac{1}{2}]$ are
\bea
\Ket{m+\textstyle{\frac{1}{2}}}_{\ell - \frac{1}{2}}
=\beta_m\Ket{m,+}-\alpha_m\Ket{m+1,-},
\ena
where $m=-\ell,-\ell+1,\ldots,\ell-1$.
\end{enumerate}
In order to diagonalize $T$ in \eqref{S2 tachyon}, 
{we have} to consider both 
{aspects of} a) and b) simultaneously,
that is, 
{we have to consider the total spin b) in a patch-wise way a).}

\paragraph{Condensation in ${\cal U}_N$}
First, we consider points in the open set ${\cal U}_N$.
As stated, $R_{N}$ in a) appears in the spin along the axis through 
$\boldsymbol{x}$.
In general, for an angular momentum operator $\boldsymbol{J}$,
the term $\boldsymbol{x} \cdot \boldsymbol{J}$ determines 
the new ``north pole'' direction through $\boldsymbol{x} $.
Then eigenvalues of $J'_3=\hat{\boldsymbol{x}} \cdot \boldsymbol{J}$ can also be 
used to label the ONB.
Here $J_i$ and $J'_3$ are related by an $SO(3)$ rotation $\Lambda^i_j$
that sends the unit vector through the point $\Omega=(\theta,\varphi)$ on the unit sphere 
to that pointing to the north pole $\boldsymbol{x}=(0,0,1)$.
This rotation is generated by the unitary operator, 
\bea
R_N(\Omega)
&=e^{-i\varphi J_3}e^{-i\theta J_2}e^{i\varphi J_3}\nn
&=e^{- \frac{1}{2}\theta  (e^{-i\varphi}J_+ -e^{i\varphi}J_-)},
\label{RNOmega}
\ena
{which} satisfies
\bea
R_N^\dagger(\Omega) J_i R_N(\Omega) =\Lambda_i^j J_j, 
\ena
with
\bea
\Lambda=
\begin{pmatrix}
\cos\varphi & -\sin\varphi & 0\\
\sin\varphi & \cos\varphi & 0\\
0 &0 & 1
\end{pmatrix}
\begin{pmatrix}
\cos\theta & 0 & \sin\theta \\
0 & 1 & 0 \\
-\sin\theta & 0 & \cos\theta
\end{pmatrix}
\begin{pmatrix}
\cos\varphi & \sin\varphi & 0\\
-\sin\varphi & \cos\varphi & 0\\
0 &0 & 1
\end{pmatrix}.
\ena
(See Appendix {\ref{appendix2}} for a proof.)
The previous $R_{N}$ in \eqref{AM2} is the spin $1/2$ case of \eqref{RNOmega}.
This implies the spin-$j$ analogue of \eqref{rotation of sigma x},
\bea
R^\dagger_N(\Omega)(\boldsymbol{x} \cdot \boldsymbol{J})R_N(\Omega)
=|\boldsymbol{x}| J_3.
\ena
In particular, the transformed state 
$R_N(\Omega)\Ket{j}$ of the highest weight state $\Ket{j}$ 
is called the Bloch (spin) coherent state {\cite{Gazeau}}.

In our case, consider \eqref{RNOmega} for the total spin 
$\boldsymbol{J}=\boldsymbol{L}+\boldsymbol{S}$.
{We can then split it} as $R_N(\Omega)=R_N^{(L)}(\Omega)R_N^{(S)}(\Omega)$, 
with \eqref{RNOmega} for $\boldsymbol{S}$ and $\boldsymbol{L}$, respectively.
{For} the tachyon profile \eqref{S2 tachyon}, 
it is obvious that this operator still diagonalizes 
$\boldsymbol{x} \cdot \boldsymbol{\sigma}=2\boldsymbol{x} \cdot \boldsymbol{S}$.
On the other hand, it keeps $\boldsymbol{\sigma} \cdot \boldsymbol{L}=2\boldsymbol{S} \cdot \boldsymbol{L}$ invariant,
since it is an $SO(3)$ scalar operator.  
Therefore, the tachyon profile is written as
\bea
T(\boldsymbol{x})
=u R_N(\Omega)(|\boldsymbol{x}|\sigma_3 -\rho \boldsymbol{\sigma} \cdot \boldsymbol{L})R^\dagger_N(\Omega).
\label{T by R}
\ena
Note that the $\Omega$=$(\theta,\varphi)$-dependence is absorbed into $R_N(\Omega)$.
It is then reasonable to use the orthonormal basis of the form 
$R_N(\Omega)\Ket{m,\epsilon}$ to find the eigenstates of $T$.
According to b), the second term $\boldsymbol{\sigma} \cdot \boldsymbol{L}$ in 
\eqref{T by R} is diagonalized by the states of the form 
$R_N(\Omega)\Ket{m+\textstyle{\frac{1}{2}}}_{\ell \pm \frac{1}{2}}$, but 
we should also take into account the first term.
It turns out that two particular states $R_N(\Omega)\Ket{\ell,+}$ and $R_N(\Omega)\Ket{-\ell,-}$ 
are already the eigenstates of $T$.
By using 
\bea
\boldsymbol{\sigma} \cdot \boldsymbol{L}
=\sigma_3 L_3 +{\textstyle \frac{1}{2}}(\sigma_+ L_- +\sigma_- L_+),
\label{famous id}
\ena
we obtain 
\bea
T R_N(\Omega)\Ket{\ell,+}
&=u R_N(\Omega)
\left\{ (|\boldsymbol{x}|-\rho L_3) \sigma_3 
-{\textstyle \frac{\rho}{2}}(\sigma_+ L_- +\sigma_- L_+)
\right\}\Ket{\ell,+}\nn
&=u 
(|\boldsymbol{x}|-\rho \ell) R_N(\Omega)\Ket{\ell,+} ,
\ena
which is zero at a point in ${\cal U}_N$ with $|\boldsymbol{x}|=\rho\ell $.
Thus, a sphere with radius $\rho\ell $
survives under the tachyon condensation.
Similarly, we have 
\bea
T R_N(\Omega)\Ket{-\ell,-}
&=u R_N(\Omega)
\left\{ (|\boldsymbol{x}|-\rho L_3) \sigma_3 
-{\textstyle \frac{\rho}{2}}(\sigma_+ L_- +\sigma_- L_+)
\right\}\Ket{-\ell,-}\nn
&=-u 
(|\boldsymbol{x}|+\rho \ell)R_N(\Omega)\Ket{-\ell,-} ,
\ena
which is always negative (no zero locus), thus,
this state is completely annihilated under the tachyon condensation.

For the remaining eigenstates, consider a two dimensional subspace of the form
\bea
R_N(\Omega) \left\{a_m \Ket{m,+} +b_m \Ket{m+1,-}\right\}
\quad \leftrightarrow
\begin{pmatrix}
a_m\\
b_m
\end{pmatrix}
\ena
for a fixed $m$ with $m=-\ell,-\ell+1,\ldots,\ell-1$ and with arbitrary coefficients 
$a_m(\boldsymbol{x})$ and $b_m(\boldsymbol{x})$.
{The point is that} $T(\boldsymbol{x})$ is closed within this subspace:
\bea
T R_N(\Omega)\Ket{m,+}
&=u R_N(\Omega)
\left\{ (|\boldsymbol{x}|-\rho L_3) \sigma_3 
-{\textstyle \frac{\rho}{2}}(\sigma_+ L_- +\sigma_- L_+)
\right\}\Ket{m,+}\nn
&=u R_N(\Omega)\left\{
(|\boldsymbol{x}|-\rho m) \Ket{m,+} 
-\rho \sqrt{(\ell-m)(\ell+m+1)}\Ket{m+1,-}\right\},\nn
T R_N(\Omega)\Ket{m+1,-}
&=u R_N(\Omega)
\left\{ (|\boldsymbol{x}|-\rho L_3) \sigma_3 
-{\textstyle \frac{\rho}{2}}(\sigma_+ L_- +\sigma_- L_+)
\right\}\Ket{m+1,-}\nn
&=u R_N(\Omega)\left\{
-(|\boldsymbol{x}|-\rho(m+1)) \Ket{m+1,-} 
-\rho \sqrt{(\ell+m+1)(\ell-m)}\Ket{m,+}\right\}.
\ena
It implies $T(\boldsymbol{x})$ is
{effectively represented as} a $2\times 2$ matrix-valued 
function $T^{(m)}(|\boldsymbol{x}|)$ for each $m$:
\bea
T^{(m)}
\begin{pmatrix}
a_m\\
b_m
\end{pmatrix}
=u
\begin{pmatrix}
|\boldsymbol{x}|-\rho m & -\rho \sqrt{(\ell-m)(\ell+m+1)}\\
-\rho \sqrt{(\ell-m)(\ell+m+1)} & -|\boldsymbol{x}|+\rho (m+1)
\end{pmatrix}
\begin{pmatrix}
a_m\\
b_m
\end{pmatrix}.
\label{def of Tm}
\ena
This matrix is diagonalized in a standard way (see Appendix {\ref{appendix3}} for more detail)
{and} the eigenvalues at each point (i.e., functions) are {found to be}
\bea
\lambda^{(m)}_{\pm}(|\boldsymbol{x}|)
&=u\left[\textstyle{\frac{\rho}{2}} \pm \left|M^{(m)}\right|\right],
\ena
with
\begin{equation}
	\left| M^{(m)} \right| \equiv
\sqrt{\rho^2(\ell-m)(\ell+m+1)
\left(|\boldsymbol{x}|-\rho (m+\textstyle{\frac{1}{2}})\right)^2}.
\end{equation}
For all $m=-\ell,-\ell+1,\ldots,\ell-1$, $|M^{(m)}|^2$ 
satisfies $|M^{(m)}|^2 \ge 2\rho^2 \ell$, and thus 
$|M^{(m)}| > \rho /2$ for all spin $\ell \ge1/2$.
This implies that for any $\ell$ and $m$, 
two eigenvalues $\lambda^{(m)}_{\pm}$ are always non-zero at any point 
$\boldsymbol{x} \in {\cal U}_N$.
Therefore, the tachyon condensation annihilates the corresponding eigenstates.

In summary, the eigenvalues of $T(\boldsymbol{x})$ and the corresponding eigenstates 
are given by 
\bea
|\boldsymbol{x}|-\rho \ell:~~&R_N(\Omega)\Ket{\ell,+},\nn
-(|\boldsymbol{x}|+\rho \ell):~~&R_N(\Omega)\Ket{-\ell,-},\nn
\lambda^{(m)}_+(|\boldsymbol{x}|) :~~&R_N(\Omega)\left\{W^{(m)}_{11}\Ket{m,+} + 
W^{(m)}_{21}\Ket{m+1,-}\right\},\nn
\lambda^{(m)}_-(|\boldsymbol{x}|) :~~&R_N(\Omega)\left\{W^{(m)}_{12}\Ket{m,+} + 
W^{(m)}_{22}\Ket{m+1,-}\right\},
\label{summary on N}
\ena
where the explicit form of the matrix $W^{(m)}(|\boldsymbol{x}|)$ 
is given in the Appendix \ref{appendix3}.
The first state becomes a zero mode at a point $\boldsymbol{x}\in {\cal U}_N$ 
with the radius $|\boldsymbol{x}|=\rho\ell$.
The other states always vanish under the tachyon condensation.

\paragraph{Condensation in ${\cal U}_S$}
In ${\cal U}_S$, the eigenvalues of the tachyon profile are the same as 
in \eqref{summary on N} in ${\cal U}_N$,
but another unitary operator $R_S$ is needed to diagonalize $T$ and 
a different state survives under the tachyon condensation.
To see this, consider a point on the negative $x^3$-axis, 
$\boldsymbol{x}=(0,0,-|\boldsymbol{x}|)$.
Because of $\boldsymbol{\sigma}\cdot \boldsymbol{x}=-|\boldsymbol{x}|\sigma_3$
and by using \eqref{famous id}, 
we find 
\bea
&T(\boldsymbol{x})\Ket{-\ell,-}=u\left(|\boldsymbol{x}|-\rho \ell \right)\Ket{-\ell,-},\nn
&T(\boldsymbol{x})\Ket{\ell,+}=u\left(-|\boldsymbol{x}|-\rho \ell \right)\Ket{\ell,+}.
\ena
It shows that the surviving state is the lowest weight state $\Ket{-\ell,-}$ around the south pole, 
if $|\boldsymbol{x}|=\rho \ell$.
This is extended to any point $\boldsymbol{x}=(|\boldsymbol{x}|,\theta,\varphi) \in {\cal U}_S$  
$(\theta \ne 0)$ by sending it  to the south pole generated by the unitary operator 
\bea
\tilde{R}_S(\Omega)
&=e^{-i\varphi J_3}e^{i(\pi-\theta)\varphi J_2}e^{i\varphi J_3}\nn
&=e^{ -\frac{1}{2}(\pi-\theta) (e^{-i\varphi}J_+ -e^{i\varphi}J_-)}.
\ena
Then, the Bloch coherent state $\tilde{R}_S(\Omega)\Ket{-\ell,-}$ with respect to the south pole
is shown to be an eigenstate of $T(\boldsymbol{x})$:
\bea
T(\boldsymbol{x})\tilde{R}_S(\Omega)\Ket{-\ell,-}
&=u\tilde{R}_S(\Omega)\left(-|\boldsymbol{x}|\sigma_3 
-\rho \boldsymbol{\sigma}\cdot\boldsymbol{L}\right)\Ket{-\ell,-}\nn
&=u\left(|\boldsymbol{x}|-\rho \ell \right)\tilde{R}_S(\Omega)\Ket{-\ell,-},
\ena
which is the zero mode if $|\boldsymbol{x}|=\rho \ell$.
Finding the other eigenstates is similar to perform.

Although this treatment is sufficient when we are interested only in ${\cal U}_S$,
it should actually be consistent with the result in ${\cal U}_N$ 
in the overlapping region ${\cal U}_{NS}={\cal U}_N \cap {\cal U}_S$.
The remaining $1$-dimensional fibers of both construction should be identified with each other,
that is, the difference should be at most a $U(1)$ phase.
To this end, we insert an extra rotation $\Theta =e^{-i\pi J_2}$ on the state in ${\cal U}_S$
to diagonalize $T(\boldsymbol{x})$.
For $\boldsymbol{J}=\boldsymbol{S}$, 
it is the Wigner time reversal operator, 
\bea
\Theta^{(S)}=e^{-i\pi \frac{\sigma_2}{2}}=-i\sigma_2=\mat{0,-1,1,0},
\ena
which flips $\Ket{+}$ and $\Ket{-}$.
In general, due to the relations
\bea
\Theta^\dagger J_{1,3}\Theta =-J_{1,3}, \quad 
\Theta^\dagger J_2 \Theta=J_2,
\ena
the state $\Theta\Ket{m,\epsilon}$ is the eigenstate with alternating the sign:
\bea
J_3 \Theta\Ket{m,\epsilon}
=-\Theta J_3\Ket{m,\epsilon}=-(m+\epsilon)\Theta\Ket{m,\epsilon}.
\ena
By the same relations, we also have
\bea
&\Theta^\dagger (\boldsymbol{\sigma}\cdot \boldsymbol{L}) \Theta 
=\boldsymbol{\sigma}\cdot \boldsymbol{L},\nn
&\Theta^\dagger (-|\boldsymbol{x}|\sigma_3) \Theta
=+|\boldsymbol{x}|\sigma_3.
\ena
We now define%
\footnote{ 
There is an constant phase ambiguity to define $\Theta$.
It can be shown that the spin $1/2$ part $R_S^{(S)}$ coincides with $R_S$ in \eqref{AM} 
so that our choice of $\Theta$ is to be consistent with {\cite{Asakawa2017}}.}
\bea
R_S(\Omega)=\tilde{R}_S(\Omega)\Theta.
\label{def of R_S}
\ena
Then the tachyon profile in ${\cal U}_S$ is written as
\bea
T(\boldsymbol{x})&=u R_S(\Omega)
 \left(|\boldsymbol{x}|\sigma_3-\rho \boldsymbol{\sigma}\cdot \boldsymbol{L}\right)
R^\dagger_S(\Omega).
\ena
Since the term inside the bracket is the same as in ${\cal U}_N$, 
it clearly shows that the eigenvalues of $T$ are the same as ${\cal U}_N$ as required, 
and the corresponding eigenstates are given by 
\bea
|\boldsymbol{x}|-\rho \ell:~~&R_S(\Omega)\Ket{\ell,+},\nn
-(|\boldsymbol{x}|+\rho \ell):~~&R_S(\Omega)\Ket{-\ell,-},\nn
\lambda^{(m)}_+ (|\boldsymbol{x}|) :~~&R_S(\Omega)\left\{W^{(m)}_{11}\Ket{m,+} + 
W^{(m)}_{21}\Ket{m+1,-}\right\},\nn
\lambda^{(m)}_- (|\boldsymbol{x}|) :~~&R_S(\Omega)\left\{W^{(m)}_{12}\Ket{m,+} + 
W^{(m)}_{22}\Ket{m+1,-}\right\}.
\ena
The first state becomes a zero mode at any point $\boldsymbol{x}\in {\cal U}_S$ 
with the radius $|\boldsymbol{x}|=\rho\ell$.
The other states always vanish under the tachyon condensation.

\paragraph{Gluing in the overlap ${\cal U}_{NS}$}
The zero mode in ${\cal U}_N$ and ${\cal U}_S$ are now written respectively as 
$R_N(\Omega)\Ket{\ell,+}$ and $R_S(\Omega)\Ket{\ell,+}$.
In the overlapping region ${\cal U}_{NS}$, they are identified 
up to a $U(1)$ gauge transformation (transition function, more properly).

To see this, it is worth to rewrite $R_S(\Omega)$ in \eqref{def of R_S} 
as (valid for $\theta \ne 0,\pi$)
\bea
R_S(\Omega)&=\tilde{R}_S(\Omega)\Theta \nn
&=e^{-i\varphi J_3} e^{i(\pi-\theta) J_2}e^{i\varphi J_3}e^{-i\pi J_2}\nn
&=(e^{-i\varphi J_3} e^{-i\theta J_2} e^{i\varphi J_3}) e^{-2i\varphi J_3} \nn
&=R_N(\Omega)e^{-2i\varphi J_3}.
\ena
By acting this on the state $\Ket{m,\epsilon}$, it implies 
\bea
R_S(\Omega) \Ket{m,\epsilon}
=e^{-2i\varphi (m+\frac{\epsilon}{2})} R_N(\Omega)\Ket{m,\epsilon}.
\ena
This shows that two states 
$R_S(\Omega) \Ket{m,\epsilon}$ and $R_N(\Omega) \Ket{m,\epsilon}$ are 
related by a $U(1)$ phase for fixed $(m,\epsilon)$.
Hence, the transition function is $U(1)^{2k}$-valued:
\bea
R^\dagger_N(\Omega)R_S(\Omega)=e^{-2i\varphi J_3}.
\ena
In particular, we obtain 
\bea
R_S(\Omega) \Ket{\ell,+}
=e^{-ik\varphi}R_N(\Omega) \Ket{\ell,+},
\ena
since $2(\ell+{\textstyle \frac{1}{2}}) =k$.
This is nothing but the $U(1)$ transition function for Wu-Yang $k$-monopole.

\paragraph{Structure of the tachyon potential}
Having found the eigenvalues of $T(\boldsymbol{x})$, it is easy to write
the tachyon potential in the spectral decomposition.
Then, it has the form \eqref{potential to delta} as
\bea
&e^{-T^2} \xrightarrow{u\to\infty} 
\frac{u}{\sqrt{\pi}}\delta(|\boldsymbol{x}| -\rho\ell) P_{N/S}(\Omega), \nn
&P_N (\Omega) 
=R_N(\Omega) \Ket{\ell,+}\Bra{\ell,+} R^\dagger_N(\Omega),\nn
&P_S (\Omega) 
=R_S(\Omega) \Ket{\ell,+}\Bra{\ell,+} R^\dagger_S(\Omega).
\label{fuzzy S^2 projection}
\ena
Here we used the fact 
$e^{-u^2 (|\boldsymbol{x}| -\rho \ell)^2} \to \frac{u}{\sqrt{\pi}}\delta(|\boldsymbol{x}| -\rho\ell)$ in the limit $u\to \infty$ on the radial delta function \cite{Asakawa2017}.

\begin{figure}[h]
\begin{center}
	\includegraphics[scale=0.7]{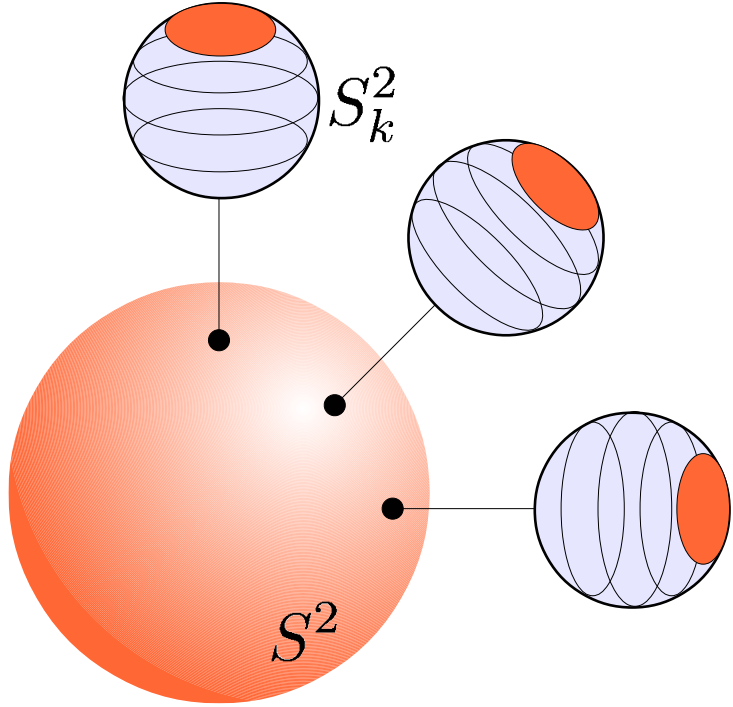}
\caption{\small 
The large sphere represents the base space $M=S^2$, and small spheres are family of 
fuzzy spheres. 
A fuzzy sphere at the north pole is divided by ring shaped regions, 
which correspond to $\Ket{m,\epsilon}$ and its top 
corresponds to the zero mode $\Ket{\ell,+}$.
If we move on the base space $M$ to the point $\Omega$, the fuzzy sphere  
is divided by regions according to the spin along $\Omega$, with its top being a coherent 
state $R_N(\Omega)\Ket{\ell,+}$.
} 
\label{fig:fuzzy}
\end{center}
\end{figure}

Matrices $\Phi$ originally give a family of fuzzy spheres on $\real^3$
as a Chan-Paton bundle of the non-BPS D3-branes.
This potential controls the reduction of both the worldvolume and the Chan-Paton space.
The worldvolume $\real^3$ of the non-BPS D3-branes reduces
to the sphere $M=S^2$ defined by $|\boldsymbol{x}| =\rho \ell$,
that is considered as a spherical D2-brane.
This sphere is commutative and embedded in $\real^3$ (thus in the spacetime as well).
At each point on the sphere specified by $\Omega=(\theta,\varphi)$, 
the original Chan-Paton space reduces to a $1$-dimensional subspace 
$R_N(\Omega)\Ket{\ell,+}$ on ${\cal U}_N$ or
$R_S(\Omega)\Ket{\ell,+}$ on ${\cal U}_S$,
that are related by a $U(1)$ transition function on ${\cal U}_{NS}$.
It is essentially the Bloch coherent state.
The schematic picture is given in Figure \ref{fig:fuzzy}.
The projection operators $P_N (\Omega)$ and $P_S (\Omega)$ also define an induced $U(1)$-gauge connection
as we will see.

It is interesting that the {obtained} radius $\rho \ell$ of $S^2$ 
is different from the expected ``radius'' of the fuzzy sphere $\rho \sqrt{\ell(\ell +1)}$.
Our result should also be compared with the radius 
given by the charge density formula \cite{Hashimoto2004}%
\footnote{The original charge density is supported on the family of spherical shells at finite $k$, but the author of \cite{Hashimoto2004} argued that the commutator corrections improve the formula to 
give a single sphere with a physical radius $\rho \sqrt{\ell(\ell +1)}$. 
Our result supports this improvement but the radius does not coincide with each other.}.
The radius can be any value at this stage, because of an arbitrary constant $\rho$.
It would be determined by the dynamics of D2-brane, 
since $\rho$ is regarded as a constant mode of a transverse scalar field on the spherical D2-brane,
as analyzed by {\cite{Myers1999}\cite{Asakawa2017}}.

\paragraph{Induced gauge connection}
The tachyon potential \eqref{fuzzy S^2 projection} defines the projective module
or equivalently a complex line bundle over $M=S^2$.
The corresponding $U(1)$ gauge connection \eqref{def of gauge field}
is given patch-wise, i.e., a gauge potentials $A_N$ on ${\cal U}_N$ 
and $A_S$ on ${\cal U}_S$ are given respectively by
\bea
&i A_N (\Omega) =\Bra{\ell,+}R^\dagger_N(\Omega) d R_N(\Omega) \Ket{\ell,+},\nn
&i A_S (\Omega) =\Bra{\ell,+}R^\dagger_S(\Omega) d R_S(\Omega) \Ket{\ell,+}.
\ena
After some algebra, we obtain 
\bea
&A_N (\Omega) ={\textstyle \frac{1}{2}}k (1-\cos\theta) d\varphi,\nn
&A_S (\Omega) =-{\textstyle \frac{1}{2}}k(1+\cos\theta) d\varphi,
\ena
where $k=2\ell+1$.
On the overlap ${\cal U}_{NS}$, 
they are related by the $U(1)$ transition function $e^{-ik\varphi}$:
\bea
A_S=e^{ik\varphi} A_N e^{-ik\varphi}-ie^{ik\varphi}de^{-ik\varphi}.
\ena
The $U(1)$ field strength is defined patch-wise by 
$F|_{{\cal U}_N}=dA_N$ and $F|_{{\cal U}_S} =dA_S$,
but in fact it is globally defined:
\bea
F=\textstyle{\frac{1}{2}}k \sin\theta d\theta\wedge d\varphi.
\ena
This configuration is nothing but the Wu-Yang $k$-monopole {\cite{Wu1976}}.
The RR-charge originally carried by $k$ D0-branes is 
maintained by this $U(1)$-flux on a spherical D2-brane,
where D0-branes are dissolved into a D2-brane.
In fact, in the Chern-Simons term for a D2-brane, 
the coupling to the RR $1$-form is 
\bea
\frac{1}{2\pi}\int_{S^2}F
=\frac{k}{4\pi}\int_{S^2}\sin\theta d\theta\wedge d\varphi
=k.
\ena
This result is independent of $\rho$.

\section{Myers term and K-homology}
\label{k-homology}

{We are considering the problem 
of mutually non-commuting matrix scalar fields $\Phi$ on multiple D0-branes.
The system of $k$ D0-branes with matrix scalar fields $\Phi$ 
reduces by the tachyon condensation to the region $M$ in 
the ($\real^3$ part of) spacetime, equipped with 
a Chan-Paton bundle $E$ over $M$ with $k$-magnetic flux.
Thus, we call the region $M$, or a pair $(M,E)$ as the shape of D0-branes.
In our simple example above, 
both $M=\real^2$ and $M=S^2$ are regarded as the worldvolume 
of a D2-brane and the resulting system $(M,E)$ is identified as D2-D0 bound state,
where $k$ D0-branes are resolved into the D2-brane.
This is consistent with that the Myers term produces D2-brane charge density.
However, it is not evident 
{if} the region $M$ can always be identified as D-branes
in a more general scalar fields $\Phi$.
In this section, we discuss the technical result in the previous section from 
conceptual grounds, and propose a possible physical interpretation of the shape 
of multiple D0-branes}%
\footnote{Discussion in this section is mainly based on the answer to the questions by S.~Terashima. We thank him for the private communication.
}.

The point of our notion of the shape is 
that it is completely independent of 
the coordinate interpretation for $\Phi$ and of the large $N$.
We only use the fact that the zero locus of the tachyon profile 
gives a defect made out of D0-branes.
The underlying belief is that all the D-brane systems are described as solitons by the tachyon condensation and that K-theory classifies all of them \cite{Sen1998,Witten1998}.
{Therefore, it is natural to understand the meaning of the shape along this belief,
instead of the coordinate interpretation.}
In the following, we elaborate on the structure of solitons and 
then propose that the shape fits nicely to the classification by the
K-homology, that is, the Poincar\'{e} dual to K-theory. 
This says that the shape is classified as a D-brane system.
In particular, 
the Myers term 
can be incorporated in K-homology.

\subsection{Structure of the solitons}

The original ABS construction 
($T(\boldsymbol{x})$ in \eqref{generic tachyon} with $\Phi=0$) 
represents a ($k$-tuple of) codimension $3$ soliton sitting at the origin.
It winds the field space $SU(2)$ once  
around $S^2$ at the asymptotic infinity $|\boldsymbol{x}| \to \infty$ in $\real^3$.
Let us first discuss to what extent the addition of the matrix scalar fields $\Phi$ on D0-branes changes the structure of the soliton from the original ABS construction of D0-branes.

One may think that adding scalar fields $\Phi$ in \eqref{generic tachyon} 
to the tachyon profile does not change 
{this asymptotic structure}
since it is just a continuous deformation of the ABS solution. 
It is true 
for finite $u$ and for finite $k$.
However, as we will soon see below, the asymptotic behavior itself can be changed 
by adding proper $\Phi$ with $k=\infty$.
Furthermore, 
even if $k$ is finite, it may affect the structure of the soliton 
in the limit of $u\to\infty$.
Note that the scalar fields $\Phi$ change the tachyon profile $T(\boldsymbol{x})$ at each point $\boldsymbol{x}$, not just at the origin.

To see this more explicitly, we first recall the Moyal case.
After the change of basis as in \eqref{displaced T}, only the zero eigen-function 
$t_{0,+}(\boldsymbol{x})=ux^3$ contributes to the remaining defect.
This zero mode has the form of a codimension $1$ kink along the $x^3$-direction.
The asymptotic behavior is $t_{0,+}(x^3=\pm \infty)=\pm \infty$, which is evidently 
different from the ABS construction before adding $\Phi$.
The kink charge is shown to be related to the D2-brane charge 
$\frac{1}{u vol(\real)} \int dx^3 \partial_3 t_{0,+}(\boldsymbol{x})=1$.
This drastic change of the asymptotic region is due to the $k\to \infty$ effect.

Next we move to the fuzzy $S^2$ case.
In this case, the asymptotic region is unchanged, 
but the structure at the origin is deformed.
After the change of basis, as seen for example in \eqref{summary on N}, 
the zero eigenfunction is 
$t_{\ell,+}(\boldsymbol{x})=u(|\boldsymbol{x}|-\rho\ell)$.
This satisfies the boundary condition 
$t_{\ell,+}(|\boldsymbol{x}| =0)=-u\rho\ell \to -\infty$
and $t_{\ell,+}(|\boldsymbol{x}| =\infty)=\infty$, that relates two different vacua.
The limit $u\to \infty$ is important in this situation.
Then, it behaves as a kink along the radial direction.
This is the same behavior {with the spherical D2-brane studied in \cite{Asakawa2017}.}
 
In both the cases, since the structure of the soliton is changed, 
it is no longer a system made of only D0-branes.
The appearance of a kink after the deformation is a sign 
of that the defect is 
actually a D2-D0 bound state.
For more general scalar fields $\Phi$, there may appear defects with all possible codimensions $0,1,2,3$.
Of course, if we start not with D3-brane but with non-BPS D9-branes, 
 all {the nine} transverse scalar fields can also be considered as a deformation of codimension nine ABS construction.

\subsection{More on the shape of D0-branes}

Although we have considered only two examples, 
the Moyal plane and the fuzzy sphere,
the analysis itself can be applied for more general cases.
In general, the shape $M$ of D0-branes 
is just the zero locus of the tachyon profile for given matrices $\Phi$.
Here an important fact is the tachyon field can always be diagonalized for any $\Phi$.
Thus, the zero locus $M$ is always determined uniquely.

When all $k\times k$ matrices $\Phi$ are diagonal, then the zero locus $M$
consists of $k$ different points in $\real^3$.
This is still true for $k=\infty$.
For example, let $\Phi^2=\hat{x}^1$ and $\Phi^2 =\hat{x}^2$ with commutative 
$[\hat{x}^1,\hat{x}^2]=0$ (i.e., $\theta=0$ in the Moyal case), then the shape
is given by the point set $M=\real^2$.
We know that this $M$ does not mean a D2-brane (Neumann boundary state along $\real^2$)
but infinitely many D0-branes aligned on $\real^2$ (a family of Dirichlet boundary states). 
Thus, $M$ itself does not see this difference.
On the other hand, in the Moyal case, we know that 
the shape $M=\real^2$ is a D2-brane worldvolume of a D2-D0 bound state, 
that is, a smooth submanifold in $\real^3$.
It is seen by noticing that a point $(z,\bar{z})$ in $M=\real^2$ and the origin is 
connected by the displacement operator $D(\alpha)$ of coherent states.
That is, the existence of differential structure is guaranteed by the unitary operator $U(z,\bar{z})$.
This is also consistent with the fact that a coordinate operator $\hat{x}^1$ of the Moyal 
plane is simultaneously a differential operator $i\theta \partial_2=[\hat{x}^1,\cdot]$.
Thus, it is important to include the information on the connection into the shape $(M,E)$,
in order to distinguish these cases.

If several zero modes appear, $M$ consists of several pieces, 
each of which may have different dimension in general.
In a very particular case, if there are $n$ zero modes and are degenerate on the same region $M$, then $E$ becomes a $U(n)$ bundle over $M$.
This is in contrast to the conventional description of D-branes.
The difference is apparent when considering fluctuations $\Phi'=\Phi +\delta \Phi$ further.
In the conventional description, $\delta \Phi$ is identified as a matrix scalar field on $M$,
but in our treatment, we seek zero locus again, and obtain another shape $(M',E')$.
In this sense, $M$ is always commutative and no matrix scalar fields appear on $M$.

Before going to K-homology, 
we make a brief comment on the 
{boundary state description of D-branes.}
A system of coincident D-branes is most rigorously defined 
by a boundary state equipped with a boundary interaction representing fields on D-branes.
In this description, D-branes have a definite position defined by a Dirichlet boundary condition,
and matrix scalar fields are treated as boundary perturbations.
A bound state of $n$ D2-branes and $k$ D0-branes can be described by either 
(1) D2-brane picture: 
D2-brane boundary state with a $U(n)$ gauge field $A$ carrying $k$ D0-brane charge, or 
(2) D0-brane picture: 
D0-brane boundary state with $U(k)$ scalar fields $\Phi$ carrying $n$ D2-brane charge.
Schematically, the equivalence of two pictures is given by
\bea
e^{-S_b[A]}\Ket{D2}=e^{-S_b[\Phi]}\Ket{D0}.
\ena
In the Moyal case, the equivalence of two pictures are shown in {\cite{Ishibashi1999}}.
Both the pictures represent the same mixed boundary condition from different viewpoints:
The picture (1) represents it as a deformation of the Neumann boundary state by 
a boundary interaction (constant $U(1)$ gauge flux)
{while} (2) does as a deformation of the Dirichlet boundary state.
{In terms of tachyon condensation, the system can also be realized
by the boundary state of non-BPS D3-branes
\bea
e^{-S_b[T]}\Ket{D3},
\label{D3 boundary state}
\ena
with a tachyon field $T$.} 
The advantage of this realization
is that both pictures are two different choices of basis for the tachyon profile and thus they are manifestly unitary equivalent.
The D0-brane picture (2) corresponds to the basis that diagonalizes the ABS construction,
and the D2-brane picture (1) corresponds to the basis that diagonalizes the full tachyon profile including $\Phi$%
\footnote{
They are analogous to the interaction and the Heisenberg picture, respectively,  
in quantum mechanics.
}.
We stress here, however, that the concept of the shape is independent of the choice of the pictures (1) and (2). Although the shape in the Moyal case happens to be well described in the picture (1), it is just by chance. For generic $\Phi$, although the boundary state \eqref{D3 boundary state} can be still defined and we can read off the shape and/or the boundary condition from this expression, there is no guarantee that the obtained shape is always well described in a specific picture like (1).

\subsection{K-homology}

The shape of D0-branes described so far
fits nicely to the classification of D-branes by the K-homology group as announced.
In particular, we emphasize that 
the Myers term can be incorporated in this classification.

Let us recall the definition of the K-homology {\cite{Douglas1982}}.
A K-cycle for a topological space $X$ is a triple $(M,E,\phi)$, 
where $M$ is a compact ${\rm spin}^c$ manifold without boundary, 
$E \to M $ is a complex vector bundle and 
$\phi:M\to X$ is a continuous map.
The (topological or geometric) K-homology group is defined by 
$K_*(X)=\{(M,E,\phi)\}/\sim$, where the equivalence relation is generated by 
(a) bordism, (b) direct sum and 
(c) vector bundle modification
defined by the relation which will be appeared in (\ref{vb modification}).
Here $\ast=0$ ($1$) corresponds to $M$ with even (odd) dimension, respectively.

Since the K-homology group is a Poincar\'e dual to the K-theory group,
it is natural to conjecture that the K-homology classifies D-branes.
This is first described in the concrete form in \cite{Asakawa2002} 
(see also previous discussions {\cite{Periwal2000,Moore2001}} 
and subsequent development {\cite{SZABO2002,Reis2006,Jia2013}}).
A K-cycle is conjectured to be a D-brane itself,
where $M$ is a worldvolume of a BPS Dp-brane,%
\footnote{More precisely, each connected component of $M$ corresponds to a worldvolume.}
$E$ is a Chan-Paton bundle on $M$,
and $\phi$ is a embedding of $M$ to the spacetime $X$.
The equivalence relations have been 
also interpreted as physical equivalences: 
(a) is a continuous deformation of a D-brane, 
(b) is a gauge symmetry enhancement of coincident D-branes, 
and (c) is a dielectric effect \cite{Asakawa2002,SZABO2002,Reis2006}.
In the following, however, we discuss that we should modify the physical interpretation of the equivalence relation (c).

We start with pointing out that there are several subtleties in
the above interpretation. 
First, since $\phi$ is not necessarily an embedding but just a continuous map,
there can be such an $M$ whose dimension is larger than that of $X$ in principle.
Therefore, precisely speaking, the physical interpretation described above 
can be applied only when we implicitly regard $\phi$ as an embedding {\cite{Baum2009}}.  
Next, 
there is no room for a matrix-valued scalar fields in K-cycles, 
since only a single scalar field ($U(1)$ part) is implicitly assumed 
when we consider $\phi$ to be the embedding. 
As a result, we cannot incorporate the Myers term in this interpretation in particular. 
The Myers term ${\rm Tr} e^{i_\Phi i_\Phi}C$ in the RR-coupling 
(Chern-Simons term) for D0-branes is originally obtained by applying T-duality
to the RR-coupling for D9-branes which includes the Chern character 
$C\wedge {\rm Tr} e^{F}$.
In the latter case, a non-trivial gauge flux $F\ne0$ is topologically distinct from $F=0$,  
indicating RR-coupling to higher rank RR-potentials, known as branes within branes \cite{Douglas:1995bn}.
This information is already incorporated as the Chern character for a 
K-cycle \cite{Douglas1982} as shown in \cite{Asakawa2002}.
Similarly, since non-commuting scalar fields $\Phi$ produce a RR-coupling 
to higher rank RR-potentials through the Myers term, 
T-duality requires that such configuration is distinguished from commuting one.
If the K-homology classifies all possible D-branes, 
it should be able to take into account the matrix scalar fields.

Now, let us turn to the situation in this paper.
In the coherent state method, 
the shape of $k$ D0-branes is a region $M$ in the spatial part $X=\real^3$ 
of the spacetime, which can naturally be identified with $\phi(M)$ in the K-cycle, 
with the canonical inclusion map $\phi$.
There is also a $U(1)$ Chan-Paton bundle with $k$-magnetic charge.
As stated, if zero locus are degenerate, 
it is extended to a non-abelian Chan-Paton bundle $\phi^* E$.
Therefore, our shape naturally corresponds to a K-cycle $(M,E,\phi)$, 
even if matrix scalar fields $\Phi$ are non-commuting.
In particular, the Myers term is implicitly incorporated in this new interpretation.

To see 
the effect of the Myers term {more explicitly}, 
we recall 
the equivalence (c), the vector bundle modification {\cite{Douglas1982}}, 
\bea
(M,E,\phi)\sim (\hat{M},\hat{H}\otimes \pi^* E, \phi \circ \pi).
\label{vb modification}
\ena
The r.h.s. is obtained from the l.h.s. through the clutching construction:
$\pi: \hat{M}\to M$ is a sphere bundle over $M$ 
whose fiber is an even dimensional sphere $S^{2n}$.
$\hat{H}\to \hat{M}$ is a complex vector bundle over $\hat{M}$ 
whose fiber is a Bott generator on $S^{2n}$.
Because of the appearance of the sphere, $\hat{M}$ has been interpreted as 
the worldvolume of a spherical D-brane \cite{Asakawa2002,SZABO2002,Reis2006}.
However, as seen by following {\cite{Douglas1982}} carefully, 
we should rather interpret 
$\hat{M}$ as a worldvolume of D$\bar{\rm D}$-system and 
$\hat{H}$ as an ABS construction representing $M$ as a codimension $2n$ soliton in 
$\real^{2n}$ (whose one point compactification is $S^{2n}$ above).
In other words, the equivalence \eqref{vb modification} should be
just a physical equivalence between a D-brane and 
the same D-brane constructed by the tachyon condensation.
This is consistent with that the image of both maps 
$\phi(M)=\phi \circ \pi(\hat{M})$ represent 
the same region in $X$ and the fiber $S^{2n}$ does not seen in $X$.

In our situation, the K-cycle $(M_0,E_0,\phi_0)$ in the l.h.s. of \eqref{vb modification}
corresponds to $k$ D0-branes without scalar fields; $\Phi=0$. 
That is, $M_0$ is a point, $\phi_0(M_0)=0$ in $X=\real^3$ and $E_0=\complex^k$
is a Chan-Paton space.
It is equivalent to the r.h.s. of \eqref{vb modification},
where $\hat{M}_0=S^4$ (the one-point compactification of $\real^4$) 
and $\hat{H}_0$ is an ABS construction 
of the codimension $4$ soliton on the D4$\bar{\rm D}$4-system.
Note that our non-BPS D3-branes are considered as a part of this system 
given by a kink solution along $x^4$-direction.
Thus, $\hat{H}_0$ here is essentially given by \eqref{generic tachyon} with $\Phi=0$.

Let us turn to the case of adding matrix scalar fields $\Phi$.
Under the present interpretation, 
the deformation of $(M_0,E_0,\phi_0)$ by $\Phi$ (with a non-zero Myers term) is naturally 
realized as the deformation of the r.h.s with the tachyon profile 
\eqref{generic tachyon} with $\Phi$.
The obtained K-cycle can be non-equivalent to the point-like K-cycle from the above argument. 
In that case, it should rather be equivalent to another K-cycle that is given by the shape of D0-branes with $\Phi$. 
In our examples, we obtain a triple $(M_1,E_1,\phi_1)$
where $M_1=S^2$ (in the Moyal case, one point compactification of $\real^2$), 
$E_1$ is a $U(1)$ Chan-Paton bundle with magnetic flux $k$, and $\phi_1:S^2\to X$.
Although the non-equivalence between $(M_0,E_0,\phi_0)$ and $(M_1,E_1,\phi_1)$ should
be proven mathematically, 
it should be consistent with the structure of solitons and the RR-coupling described above.

In summary, we propose that the shape of D0-branes with scalar fields 
corresponds to a K-cycle.
We claim that the Myers term (non-commuting scalar fields) is incorporated as a non-equivalent deformation of K-cycles rather than the vector bundle modification.

\section{Conclusion and discussion}
\label{sec:comments}

%

We considered D-brane systems with non-commuting scalar fields $\Phi$ via tachyon condensation and gave a novel prescription to read off the shape of the noncommutative D-brane system {as a commutative region in spacetime,} 
by {rearranging} the idea of the method proposed in \cite{Berenstein2012,Ishiki2015,Schneiderbauer2016}
(the coherent state method) {as the} tachyon condensation. 
In this interpretation, the shape of D-brane is defined as a set of zeros of the tachyon field
{together with a gauge flux on it}. 
As typical examples, we closely investigated the Moyal plane and the fuzzy sphere but the generalization to other systems is straightforward.
The point is that diagonalizing a tachyon profile is always
possible for any matrix valued scalar fields $\Phi$.
We also 
{argued that the shapes} fit well to the classification of D-branes by the K-homology group.
{This shows that the D-branes made through the Myers term are incorporated in this classification.}

{Since we focused mainly on the topological aspects of the shape $M$ corresponding to the K-homology, there are several issues that we did not touch upon.
In this section, we briefly discuss two other aspects of the shape.}

\paragraph{{Metric on the shape}}


From the point of view of the coherent state method, it is natural to define a metric 
{of the shape} only from the matrices $\Phi$ {\cite{Ishiki2015,Ishiki2016}. 
In the present context, it is suitable to be defined on the zero mode of the Chan-Paton bundle.}  
There are several notions of metrics defined on a family of Hilbert spaces, such as the quantum Fisher metric, the Fubini-Study metric and the fidelity susceptibility.
Here we adopt the definition of \cite{Provost1980,Zanardi2007}. 

Let $\Ket{\psi (q)}$ be a state depending on external parameters denoted by $q^i$.
In the information theoretic geometry, the metric on the parameter space is defined by
\bea
g_{ij} ={\rm Re} (C_{ij} -A_i A_j) ,
\label{def of info metric}
\ena
where the quantities $C_{ij}$ and $A_i$ are
\bea
&C_{ij}(q) dq^i dq^j =|\psi (q+dq)-\psi(q)|^2 =\Bracket{\p_i \psi (q)}{\p_j \psi (q)}dq^i dq^j,\nn
&A_i (q)= -i \Bracket{\psi(q)}{\p_i \psi (q)}.
\label{def of C and A}
\ena
In our case, a tachyon zero mode has the form 
$\Ket{\psi_0(\boldsymbol{x})}=U(\boldsymbol{x})\Ket{0}$, where 
$\boldsymbol{x} \in \real^3$ is considered to be parameters 
and $U(\boldsymbol{x})$ is determined by $\Phi$.
Then, the above quantities are written as
\bea
&C_{ij}(\boldsymbol{x}) =\Bra{0}\p_i U(\boldsymbol{x})^\dagger \p_j U(\boldsymbol{x})\Ket{0},\quad 
A_i (\boldsymbol{x})= -i \Bra{0}U(\boldsymbol{x})^\dagger \p_i U(\boldsymbol{x}) \Ket{0}.
\label{C and A}
\ena
The latter is nothing but the induced gauge potential \eqref{def of gauge field}.
This metric gives a length between zero mode states $\Ket{\psi_0(\boldsymbol{x})}$
and $\Ket{\psi_0(\boldsymbol{x}+d\boldsymbol{x})}$ at two nearby points%
\footnote{Of course, they should belong to the same open set in $M$.}
essentially through the overlap $\Bracket{\psi_0(\boldsymbol{x}+d\boldsymbol{x})}{\psi_0(\boldsymbol{x})}$.
In other words, this length is defined along the fiber direction of the Chan-Paton bundle.

Applying it to the Moyal case, we obtain {(see Appendix \ref{appendix1} for a proof)}
\bea
&C_{ij}(z,\bar{z}) 
=\Bra{0,+}\p_i U(z)^\dagger \p_j U(z)\Ket{0,+}
=\Bra{0}\p_i D(\alpha)^\dagger \p_j D(\alpha)\Ket{0},\\
&A_i (z,\bar{z})= -i \Bra{0,+}U(z)^\dagger \p_i U(z) \Ket{0,+}
= -i \Bra{0}D(\alpha)^\dagger \p_i D(\alpha) \Ket{0},
\label{C and A for Moyal}
\ena
and the metric becomes
\bea
ds^2=d\alpha d\bar{\alpha}
=\frac{1}{2\theta}dz d\bar{z}. 
\label{info met flat}
\ena
This is a flat metric on $\real^2$ but different from 
the induced metric of the flat Euclidean background by {a} Weyl factor.

For the fuzzy sphere case, 
the quantities in \eqref{C and A} should be evaluated 
patch-wise with respect to the state
$R_N (\Omega)\Ket{\ell,+}$ on ${\cal U}_N$ and 
$R_S (\Omega)\Ket{\ell,+}$ on ${\cal U}_S$, respectively.
It turns out, however, that the line element is the same in both 
of ${\cal U}_N$ and ${\cal U}_S$ {(see Appendix \ref{appendix4})}.
\bea
ds^2=\frac{1}{2}\left(\ell+\frac{1}{2}\right)
\left( d\theta^2 +\sin^2 \theta d\varphi^2 \right)
=\frac{k}{4}d^2\Omega.
\label{info met FS}
\ena
This is the round metric on $S^2$ with a Weyl factor.
This metric should be compared with two alternative metrics:
the induced metric {of} the sphere with radius $|\boldsymbol{x}|=\rho\ell$, 
$ds^2=\rho^2\ell^2 d^2\Omega $, {and the metric of the}
fuzzy sphere with radius $\rho \sqrt{\ell(\ell+1)}$, 
$ds^2=\rho^2 \ell(\ell+1)d^2\Omega$.
Again, the information metric (\ref{info met FS}) is different from both of them 
by {a} Weyl factor. 

Although the difference between the 
information metric and the induced metric of the flat target space 
is only the Weyl factors in these examples,
this is not the case in general. 
This can be most easily checked by adding perturbations to $\Phi$ in 
both examples. 
This difference can also be intuitively understood as follows.
In the above examples, 
the Weyl factors are given by the inverse of the noncommutative parameters,
which are also related to the densities of the D0-branes.
The induced metric just depends on the shape in the target space, while
the information metric picks up information of the density distribution 
of D0-branes.
At least in the large-$N$ limit, there exist two 
configurations of D0-branes such that they 
have a common shape in the target space but have 
different density distributions. 
Such two configurations will share the same induced metric but 
have different information metrics. This implies the inequivalence of 
the two metrics.

The appearance of the noncommutative parameters in the information metric 
also suggests that the information metric is the K\"{a}hler metric associated
with the symplectic structure given by the gauge flux.
In the above examples, the information metrics are indeed the K\"{a}hler 
metrics. 
In \cite{Ishiki2016}, it is also shown for a wide class of matrix 
configurations that the information metric 
is indeed reduced to the K\"{a}hler metric in the large-$N$ limit.

\paragraph{Effective theory on the shape}

We close this paper with rather speculative discussion.
We come back to the example of the fuzzy sphere.
The shape $(M,E)$ in this case is given by $M=S^2$ and $E$ is a complex line 
bundle over $M$ equipped with the $k$-monopole connection.
This connection comes from displacements of the Bloch coherent states 
and this suggests that a smooth structure is guaranteed to exist.
However, the situation is different from the Moyal case, 
because the fuzzy sphere is made of finite matrices.
This is easily seen by considering algebra of functions on a fuzzy sphere and 
a commutative sphere.
The algebra of functions on a fuzzy sphere is called the fuzzy spherical harmonics,
{which} corresponds to the ordinary spherical harmonics with 
{a} restriction {in} the maximal 
angular momentum in order to match the degrees of freedom%
\footnote{In our case, because a monopole exists, it is better to think about (fuzzy) 
monopole harmonics.}.
In order to close the latter algebra by restricted harmonics, it is needed to deform 
the product to a noncommutative one ($*$-product).
Although our shape $S^2$ is a commutative region in spacetime, 
when considering functions on it, 
this suggests that it behaves as a noncommutative space.

This is not a contradiction because the function algebra is needed 
only if we consider an effective field theory on the shape.
Of course, we do not need to consider a fluctuation 
as transverse scalar fields on the shape as stated before.
When the fuzzy sphere configuration $\Phi$ corresponds to the shape $(M=S^2,E)$,
then adding fluctuations $\Phi'=\Phi+\delta\Phi$ gives another shape $(M',E')$.
However, it would also be convenient to find the effective theory description,
as in the conventional D2-brane picture.
That is, the shape is kept as $(M,E)$ but $\delta\Phi$ is treated as a field on the shape.

To find a new shape caused by a small fluctuation, 
the standard perturbation theory in quantum mechanics can be applied. 
The perturbed tachyon profile $T[\Phi']$ can be considered as a Dirac-like operator with an interaction term $u\boldsymbol{\sigma}\cdot\delta\Phi$. 
Then, the new zero mode of $T[\Phi']$ will be given by 
a linear combination of the ONB for unperturbed $T[\Phi]$. 
For this purpose, the complete set of ONB found in this paper can be used. 

In the language of boundary state, this procedure is understood as follows.
The boundary state of the D0-brane picture is $e^{-S_b[\Phi']}\Ket{D0}$ 
with scalar fields $\Phi'=\Phi+\delta\Phi$. 
By realizing it as the D3-brane boundary state \eqref{D3 boundary state}, 
it would be rewritten as the form $e^{-S_b[\delta \Phi]}\Ket{M,E}$.
Here the state $\Ket{M,E}$ corresponds to the shape $M=S^2$ for a fuzzy sphere.
It would not be the conventional Neumann boundary state along $S^2$ direction,
but will be the variant of the mixed boundary state, if the shape behaves as 
a noncommutative space. 
The fields on the shape $S^2$ is 
extracted by the boundary interaction $e^{-S_b[\delta \Phi]}$.
It is interesting to see whether the effective theory for $\delta \Phi$ is given by a noncommutative field theory on $M=S^2$ with a $*$-product.
This problem is closely related to the situation of the Seiberg-Witten map \cite{Seiberg1999}.
It is interesting to study fluctuations around the Moyal plane and the fuzzy sphere 
and investigate the relation to the Seiberg-Witten map.
We hope to report on this issue in the near future.

%

\section*{Acknowledgements}
We would like to thank 
S.~Terashima 
for useful discussions and comments.
The work of G.~I.~ was supported, in part, 
by Program to Disseminate Tenure Tracking System, 
MEXT, Japan and by KAKENHI (16K17679). 
The work of S.~M.~  was supported in part by 
Grant-in-Aid for Scientific Research (C) 15K05060. 

\appendix

\section{Computational details}

\subsection{Gauge flux and metric for Moyal case}
\label{appendix1}
For the displacement operator \eqref{displacement op}, we first show relations
\bea
&\p_\alpha D(\alpha) =D(\alpha) (\had +{\textstyle \frac{\bar{\alpha}}{2}}),
&\p_{\bar{\alpha}}D(\alpha) =-D(\alpha) (\ha +{\textstyle \frac{\alpha}{2}}),\label{relations1}\\
&\p_\alpha D^\dagger (\alpha) 
=-(\had +{\textstyle \frac{\bar{\alpha}}{2}})D^\dagger (\alpha) ,
&\p_{\bar{\alpha}}D^\dagger (\alpha) 
=(\ha +{\textstyle \frac{\alpha}{2}})D^\dagger (\alpha) ,\label{relations2}
\ena
and then calculate a gauge potential and a metric.
To this end, we will use an identity 
\bea
\frac{d}{dt}e^{B(t)}=e^{B(t)}d\exp_{-B(t)} (B'(t)),
\label{op identity}
\ena
which is valid for any operator $B(t)$ with a parameter $t$.
Here $B'(t)=\frac{d}{dt}B(t)$ and 
\bea
d\exp_{B} (C) =\sum_{l=0}^\infty \frac{1}{(l+1)!}({\rm ad}_{B})^l(C)
=\frac{e^{{\rm ad}_B}-{\rm id.}}{{\rm ad}_B}(C).
\label{dexp}
\ena
First we set $B(\alpha)=\alpha \had-\bar{\alpha}\ha$.
Then, we have $\p_\alpha B(\alpha)=\had$ and 
\bea
{\rm ad}_{-B} (\p_\alpha B)
&=[-\alpha \had+\bar{\alpha}\ha, \had]
=\bar{\alpha}.
\ena
The higher order terms $({\rm ad}_{-B})^l (\p_\alpha B)$ ($l\ge2$) vanish
so that \eqref{dexp} becomes
\bea
d\exp_{-B} (\p_\alpha B)
&=\had+\textstyle{\frac{1}{2}}\bar{\alpha},
\ena
and we obtain 
\bea
\p_\alpha D(\alpha) =D(\alpha) (\had+\textstyle{\frac{1}{2}}\bar{\alpha}).
\ena
Similarly, by setting $B (\bar{\alpha})=\alpha \had-\bar{\alpha}\ha$,
we have $\p_{\bar{\alpha}} B=-\ha$ and 
\bea
&{\rm ad}_{-B} (\p_{\bar{\alpha}}B)
=[-\alpha \had+\bar{\alpha}\ha, -\ha]
=-\alpha, \nn
\Rightarrow ~~&d\exp_{-B} (\p_{\bar{\alpha}}B)
=-\ha-\textstyle{\frac{1}{2}}\alpha, \nn
\Rightarrow ~~&\p_{\bar{\alpha}} D(\alpha) 
=-D(\alpha) (\ha+\textstyle{\frac{1}{2}}\alpha).
\ena
The others in \eqref{relations2} are obtained by $\p D^\dagger =-D^\dagger \p D D^\dagger $.
For the gauge potential, because of $\alpha=z/\sqrt{2\theta}$, we needs to estimate
\bea
dD(\alpha)
&=dz\p_z D(\alpha) + d\bar{z}\p_{\bar{z}}D(\alpha) \nn
&=d\alpha \p_\alpha D(\alpha) + d\bar{\alpha}\p_{\bar{\alpha}}D(\alpha).
\ena
By using \eqref{relations1}, we obtain 
\bea
A_\alpha 
&=-i \Bra{0}D^\dagger (\alpha) \p_\alpha D (\alpha)\Ket{0}
=-i \Bra{0}\had +{\textstyle \frac{\bar{\alpha}}{2}}\Ket{0}
=-i{\textstyle \frac{\bar{\alpha}}{2}},\nn
A_{\bar{\alpha}}
&=-i \Bra{0}D^\dagger (\alpha) \p_{\bar{\alpha}} D (\alpha)\Ket{0}
=i \Bra{0}\ha +{\textstyle \frac{\alpha}{2}}\Ket{0}
=i{\textstyle \frac{\alpha}{2}},
\label{A for Moyal}
\ena
and thus 
\bea
A=A_\alpha d\alpha +A_{\bar{\alpha}}d\bar{\alpha}
=-\frac{i}{2}(\bar{\alpha}d\alpha -\alpha d\bar{\alpha})
=-\frac{i}{4\theta}(\bar{z}dz -z d\bar{z}).
\ena
Next we calculate the metric.
In \eqref{C and A}, $A_i$ is given by \eqref{A for Moyal} and $C_{ij}$ is obtained as 
\bea
C_{\alpha\alpha}
&=\Bra{0}\p_\alpha D^\dagger (\alpha) \p_\alpha D (\alpha)\Ket{0}
=-\Bra{0} (\had +{\textstyle \frac{\bar{\alpha}}{2}})^2\Ket{0}
=-{\textstyle \frac{\bar{\alpha}^2}{4}},\nn
C_{\bar{\alpha}\bar{\alpha}}
&=\Bra{0}\p_{\bar{\alpha}} D^\dagger (\alpha) \p_{\bar{\alpha}} D (\alpha)\Ket{0}
=-\Bra{0} (\ha +{\textstyle \frac{\alpha}{2}})^2\Ket{0}
=-{\textstyle \frac{\alpha^2}{4}},\nn
C_{\alpha\bar{\alpha}}
&=\Bra{0}\p_\alpha D^\dagger (\alpha) \p_{\bar{\alpha}} D (\alpha)\Ket{0}
=\Bra{0} (\ha +{\textstyle \frac{\alpha}{2}})(\had +{\textstyle \frac{\bar{\alpha}}{2}})\Ket{0}
={\textstyle \frac{|\alpha|^2}{4}}+1,\nn
C_{\bar{\alpha}\alpha}
&=\Bra{0}\p_{\bar{\alpha}} D^\dagger (\alpha) \p_{\alpha} D (\alpha)\Ket{0}
=\Bra{0} (\had +{\textstyle \frac{\bar{\alpha}}{2}})(\ha +{\textstyle \frac{\alpha}{2}})\Ket{0}
={\textstyle \frac{|\alpha|^2}{4}}.
\ena
By using these, the components in the metric are
\bea
g_{\alpha\alpha}
&=-{\textstyle \frac{\bar{\alpha}^2}{4}}+{\textstyle \frac{\bar{\alpha}^2}{4}}=0,\nn
g_{\bar{\alpha}\bar{\alpha}}
&=-{\textstyle \frac{\alpha^2}{4}}+{\textstyle \frac{\alpha^2}{4}}=0,\nn
g_{\alpha\bar{\alpha}}
&={\textstyle \frac{|\alpha|^2}{4}}+1-{\textstyle \frac{|\alpha|^2}{4}}=1,\nn
g_{\bar{\alpha}\alpha}
&={\textstyle \frac{|\alpha|^2}{4}}-{\textstyle \frac{|\alpha|^2}{4}}=0,
\ena
and thus the line element becomes
\bea
ds^2=d\alpha d\bar{\alpha}.
\ena

\subsection{Rotation}
\label{appendix2}
Let $\Lambda$ be the rotation matrix that sends 
$\boldsymbol{x}=(x^1,x^2,x^3)$ to $\boldsymbol{x}=(0,0,r)$, and let $R$ be the corresponding 
unitary operator such that 
\bea
\Lambda^i_j J^j =R^\dagger J_i R.
\ena
We have two possibilities:
\begin{enumerate}
\item[(a)] Rotation about an axis $\boldsymbol{n}=(-\sin\varphi,\cos\varphi,0)$ with an angle $-\theta$.
\item[(b)] The sequence of (1) rotation about an axis $\boldsymbol{n}=(0,0,1)$ with an angle $-\varphi$, 
(2) rotation about an axis $\boldsymbol{n}=(0,1,0)$ with an angle $-\theta$, and 
(3) rotation about an axis $\boldsymbol{n}=(0,0,1)$ with an angle $\varphi$.
\end{enumerate}
We will see (b) first.
The operation (1) is generated by $R_1=e^{i(-\varphi)J_3}=e^{-i\varphi J_3}$.
In fact,
\bea
&R^\dagger_1 J_1 R_1=e^{i\varphi J_3}J_1 e^{-i\varphi J_3} =\cos\varphi J_1 -\sin\varphi J_2,\nn
&R^\dagger_1 J_2 R_1=e^{i\varphi J_3}J_2 e^{-i\varphi J_3} =\cos\varphi J_2 +\sin\varphi J_1,\nn
&R^\dagger_1 J_3 R_1=e^{i\varphi J_3}J_3 e^{-i\varphi J_3} =J_3,
\ena
which means 
\bea
\Lambda_1=\begin{pmatrix}
\cos\varphi & -\sin\varphi & 0\\
\sin\varphi & \cos\varphi & 0\\
0 &0 & 1
\end{pmatrix}.
\ena
The operation (2) is generated by $R_2=e^{i(-\theta)J_2}=e^{-i\theta J_2}$.
In fact,
\bea
&R_2^\dagger J_1 R_2=e^{i\theta J_2}J_1 e^{-i\theta J_2} =\cos\theta J_1 +\sin\theta J_3,\nn
&R_2^\dagger J_2 R_2=e^{i\theta J_2}J_2 e^{-i\theta J_2} =J_2, \nn
&R_2^\dagger J_3 R_2=e^{i\theta J_2}J_3 e^{-i\theta J_2} =\cos\theta J_3 -\sin\theta J_1,
\ena
which means 
\bea
\Lambda_2=\begin{pmatrix}
\cos\theta & 0 & \sin\theta \\
0 & 1 & 0 \\
-\sin\theta & 0 & \cos\theta
\end{pmatrix}.
\ena
The operation (3) is generated by $R_3=e^{i\varphi J_3}$.
In fact,
\bea
&R_3^\dagger J_1 R_3=e^{-i\varphi J_3}J_1 e^{i\varphi J_3} =\cos\varphi J_1 +\sin\varphi J_2,\nn
&R_3^\dagger J_2 R_3=e^{-i\varphi J_3}J_2 e^{i\varphi J_3} =\cos\varphi J_2 -\sin\varphi J_1,\nn
&R_3^\dagger J_3 R_3=e^{-i\varphi J_3}J_3 e^{i\varphi J_3} =J_3,
\ena
which means 
\bea
\Lambda_3=\begin{pmatrix}
\cos\varphi & \sin\varphi & 0\\
-\sin\varphi & \cos\varphi & 0\\
0 &0 & 1
\end{pmatrix}.
\ena
Then the sequence of (1) to (3) is generated by $R=R_1 R_2 R_3$ and 
\bea
\Lambda=\Lambda_1 \Lambda_2 \Lambda_3 =
\begin{pmatrix}
\cos\varphi & -\sin\varphi & 0\\
\sin\varphi & \cos\varphi & 0\\
0 &0 & 1
\end{pmatrix}
\begin{pmatrix}
\cos\theta & 0 & \sin\theta \\
0 & 1 & 0 \\
-\sin\theta & 0 & \cos\theta
\end{pmatrix}
\begin{pmatrix}
\cos\varphi & \sin\varphi & 0\\
-\sin\varphi & \cos\varphi & 0\\
0 &0 & 1
\end{pmatrix}.
\ena
On the other hand, $R$ is rewritten as 
\bea
R&=R_1 R_2 R_3
=e^{-i\varphi J_3}e^{-i\theta J_2}e^{i\varphi J_3}\nn
&=\exp (-i\theta e^{-i\varphi J_3}J_2 e^{i\varphi J_3})
=e^{-i\theta (\cos\varphi J_2 -\sin\varphi J_1)}
=e^{i(-\theta) (-\sin\varphi J_1 +\cos\varphi J_2 )},
\ena
which says that $R$ generates (a).
By using $J_\pm=J_1\pm iJ_2$, $R$ is also written as 
\bea
R=e^{ -\frac{1}{2}\theta  (e^{-i\varphi}J_+ -e^{i\varphi}J_-)}.
\ena

\paragraph{The case of spin $\frac{1}{2}$}
The unitary operator $R$ in this case is given by
\bea
R&=e^{-i\varphi S_3}e^{-i\theta S_2}e^{i\varphi S_3}
=e^{-i\frac{\varphi}{2} \sigma_3}e^{-i\frac{\theta}{2} \sigma_2}
e^{i\frac{\varphi}{2} \sigma_3}\nn
&=\mat{e^{-i\frac{\varphi}{2}},0,0,e^{i\frac{\varphi}{2}}}
\mat{\cos \frac{\theta}{2}, -\sin\frac{\theta}{2},\sin\frac{\theta}{2},\cos\frac{\theta}{2}}
\mat{e^{i\frac{\varphi}{2}},0,0,e^{-i\frac{\varphi}{2}}}\nn
&=\mat{\cos \frac{\theta}{2}, -\sin\frac{\theta}{2}e^{-i\varphi},
\sin\frac{\theta}{2}e^{i\varphi},\cos\frac{\theta}{2}}.
\ena

\subsection{Details on the diagonalization}
\label{appendix3}
In general, a $2\times 2$ matrix of the form $M=M_0\mathbf{1}_2 +M_i \sigma^i$ 
has eigenvalues $\lambda_\pm=M_0\pm |M|$, and is diagonalized either by 
\bea
W_1=\frac{1}{\sqrt{2|M|(|M|+M_3)}}\mat{|M|+M_3,-M_1+iM_2,M_1+iM_2,|M|+M_3},
\label{W1}
\ena
if $|M|+M_3\ne0 $, or
\bea
W_2=\frac{1}{\sqrt{2|M|(|M|-M_3)}}\mat{M_1-iM_2,|M|-M_3,|M|-M_3,-M_1-iM_2},
\label{W2}
\ena
if $|M|-M_3\ne0 $, where $|M|=\sqrt{M_i M^i}$.
That is, $M$ is written as
\bea
M=W_{1,2} \mat{\lambda_+,0,0,\lambda_-} W^\dagger_{1,2}.
\ena
The eigenstates $v_\pm$ with eigenvalues $\lambda_\pm$ 
are given by two column vectors in $W_1$ (and similar for $W_2$):
\bea
v_+=\frac{1}{\sqrt{2|M|(|M|+M_3)}}
\begin{pmatrix}
|M|+M_3\\
M_1+iM_2
\end{pmatrix}, \quad
v_-=\frac{1}{\sqrt{2|M|(|M|+M_3)}}
\begin{pmatrix}
-M_1+iM_2\\
|M|+M_3
\end{pmatrix}.
\ena

In our case, $T^{(m)}$ in \eqref{def of Tm} is written in this form by
\bea
&T^{(m)}=u(M^{(m)}_0\mathbf{1}_2 +M^{(m)}_i \sigma^i) ,\nn
&M^{(m)}_0=\textstyle{\frac{\rho}{2}}, \quad 
M^{(m)}_1=-\rho\sqrt{(\ell-m)(\ell+m+1)},\quad  
M^{(m)}_2=0, \quad 
M^{(m)}_3=|\boldsymbol{x}|-\rho (m+\textstyle{\frac{1}{2}}).
\label{decomposition of Tm}
\ena
Then, eigenvalues $\lambda^{(m)}_{\pm}$ of $T^{(m)}$ are
\bea
\lambda^{(m)}_{\pm}(|\boldsymbol{x}|)
&=u(M^{(m)}_0\pm |M^{(m)}|)\nn
&=u\left[\textstyle{\frac{\rho}{2}} 
\pm \sqrt{\rho^2(\ell-m)(\ell+m+1)
+\left(|\boldsymbol{x}|-\rho (m+\textstyle{\frac{1}{2}})\right)^2} \right],
\ena
where we have used
\bea
|M^{(m)}|^2 =M^{(m)}_i M^{(m)i} 
&=\rho^2(\ell-m)(\ell+m+1)
+\left(|\boldsymbol{x}|-\rho (m+\textstyle{\frac{1}{2}})\right)^2.
\ena
Note that it is also written as
\bea
M^{(m)}_i M^{(m)i}
&=|\boldsymbol{x}|^2 -2\rho |\boldsymbol{x}|  \left(m+\textstyle{\frac{1}{2}}\right)
+\rho^2 \left(\ell+\textstyle{\frac{1}{2}}\right)^2.
\ena

Next, we will check whether $W_{1,2}$ in \eqref{W1} and \eqref{W2} are allowed.
Because $|M^{(m)}|^2=(M^{(m)}_1)^2+(M^{(m)}_3)^2$, we have 
\bea
|M^{(m)}| =M^{(m)}_3
& \Leftrightarrow |M^{(m)}|^2 =(M^{(m)}_3)^2 ~\text{and}~ M^{(m)}_3>0 \nn
& \Leftrightarrow (M^{(m)}_1)^2 =0 ~\text{and}~ M^{(m)}_3>0,\nn
|M^{(m)}| =-M^{(m)}_3
& \Leftrightarrow |M^{(m)}|^2 =(M^{(m)}_3)^2 ~\text{and}~ M^{(m)}_3<0 \nn
& \Leftrightarrow (M^{(m)}_1)^2 =0 ~\text{and}~ M^{(m)}_3<0.
\ena
Since $(M^{(m)}_1)^2 \ne 0$ in our case, both $W_1$ and $W_2$ are allowed.
Note that in our definition \eqref{decomposition of Tm}, $u$ is extracted, but 
\eqref{W1} and \eqref{W2} are still correct and are $u$-independent.
We choose $W_1$, that is, 
\bea
&T^{(m)}=W^{(m)} \mat{\lambda^{(m)}_+, 0,0, \lambda^{(m)}_-}W^{(m)\dagger},\nn
\quad
&W^{(m)}=\mat{W^{(m)}_{11},W^{(m)}_{12},W^{(m)}_{21},W^{(m)}_{22}}
=\frac{1}{\sqrt{C^{(m)}}}
\mat{|M^{(m)}|+M^{(m)}_3,-M^{(m)}_1,M^{(m)}_1,|M^{(m)}|+M^{(m)}_3},
\ena
where we define $C^{(m)}=2|M^{(m)}|(|M^{(m)}|+M^{(m)}_3)$.
As an operator, $W^{(m)}$ is written as
\bea
W^{(m)}=&
W^{(m)}_{11}\Ket{m,+}\Bra{m,+}
+W^{(m)}_{12}\Ket{m+1,-}\Bra{m,+} \nn
&+W^{(m)}_{21}\Ket{m,+} \Bra{m+1,-}
+W^{(m)}_{22}\Ket{m+1,-} \Bra{m+1,-}.
\ena
Then two eigenvalues of $T$ and the corresponding eigenstates are given by
\bea
\lambda^{(m)}_+ :~~&R_N(\Omega)\left\{W^{(m)}_{11}\Ket{m,+} + 
W^{(m)}_{21}\Ket{m+1,-}\right\},\nn
\lambda^{(m)}_- :~~&R_N(\Omega)\left\{W^{(m)}_{12}\Ket{m,+} + 
W^{(m)}_{22}\Ket{m+1,-}\right\}.
\ena
For the later purpose, we define a unitary operator $W_N$,
which acts as $W^{(m)}$ on each subspace
${\rm span}\{\Ket{m,+},\Ket{m+1,-}\}$ for $m$,
and $1$ for $\Ket{\ell,+}$ and $\Ket{-\ell,-}$:
\bea
W_N(|\boldsymbol{x}|)
&=\Ket{\ell,+}\Bra{\ell,+}+\Ket{-\ell,-}\Bra{-\ell,-}
+\sum_{m=-\ell}^{\ell-1}W^{(m)}.
\ena
It depends on $|\boldsymbol{x}|$ but is independent of $\Omega$.
Then, the tachyon field is written as
\bea
T(\boldsymbol{x})
=W_N(|\boldsymbol{x}|) R_N(\Omega)
\Lambda (|\boldsymbol{x}|)
R^\dagger_N(\Omega)W^\dagger_N (|\boldsymbol{x}|),
\ena
where $\Lambda$ denotes an hermitian operator of eigenvalues
\bea
\Lambda (|\boldsymbol{x}|)
&=u(|\boldsymbol{x}|-\rho \ell)\Ket{\ell,+}\Bra{\ell,+}
-u(|\boldsymbol{x}|+\rho \ell)\Ket{-\ell,-}\Bra{-\ell,-}\nn
&+\sum_{m=-\ell}^{\ell-1}\lambda^{(m)}_+ \Ket{m,+}\Bra{m,+}
+\lambda^{(m)}_- \Ket{m+1,-}\Bra{m+1,-}.
\ena

\subsection{Gauge flux and metric for fuzzy sphere case}
\label{appendix4}
\paragraph{In the open set ${\cal U}_N$}
For \eqref{RNOmega}, we first show the relations,
\bea
&\p_\theta R_N(\Omega ) 
=R_N(\Omega )\textstyle{\frac{1}{2}}(e^{i\varphi}J_- -e^{-i\varphi}J_+),\label{RNid1}\\
&\p_\theta R^\dagger_N(\Omega ) 
=-\textstyle{\frac{1}{2}}(e^{i\varphi}J_- -e^{-i\varphi}J_+)R^\dagger_N(\Omega ),\label{RNid2}\\
&\p_\varphi R_N(\Omega ) 
=R_N(\Omega ) \left[
i(1-\cos\theta)J_3 +\textstyle{\frac{i}{2}}\sin\theta (e^{i\varphi}J_- +e^{-i\varphi}J_+)
\right],\label{RNid3}\\
&\p_\varphi R^\dagger_N(\Omega ) 
=-\left[
i(1-\cos\theta)J_3 +\textstyle{\frac{i}{2}}\sin\theta (e^{i\varphi}J_- +e^{-i\varphi}J_+)
\right]R^\dagger_N(\Omega ) .\label{RNid4}
\ena
\eqref{RNid2} and \eqref{RNid4} follow from \eqref{RNid1} and \eqref{RNid3}, respectively.
To show \eqref{RNid1} and \eqref{RNid3}, we will use the identity \eqref{op identity} again.
By setting $B(\theta,\varphi)=\frac{1}{2}\theta (e^{i\varphi}J_- -e^{-i\varphi}J_+)$, 
we have
\bea
\p_\theta B(\theta,\varphi) 
&=\textstyle{\frac{1}{2}}(e^{i\varphi}J_- -e^{-i\varphi}J_+), \nn
\p_\varphi B(\theta,\varphi) 
&=\textstyle{\frac{i}{2}}\theta (e^{i\varphi}J_- +e^{-i\varphi}J_+).
\label{pB}
\ena
From the first line of \eqref{pB}, it is obvious that 
\bea
{\rm ad}_{-B} (\p_\theta B)
&=[-\textstyle{\frac{1}{2}}\theta (e^{i\varphi}J_- -e^{-i\varphi}J_+),
\textstyle{\frac{1}{2}}(e^{i\varphi}J_- -e^{-i\varphi}J_+)]=0,
\ena
and that the higher order terms vanish.
Thus, we find
\bea
&R^\dagger_N(\Omega) \p_\theta R_N(\Omega )
=d\exp_{-B} (\p_\theta B)
=\textstyle{\frac{1}{2}}(e^{i\varphi}J_- -e^{-i\varphi}J_+),
\label{key relation for RdR1}
\ena
which leads to \eqref{RNid1}.
From the second line of \eqref{pB}, we find 
\bea
{\rm ad}_{-B} (\p_\varphi B)
&=[-\textstyle{\frac{1}{2}}\theta (e^{i\varphi}J_- -e^{-i\varphi}J_+),
\textstyle{\frac{i}{2}}\theta (e^{i\varphi}J_- +e^{-i\varphi}J_+)]\nn
&=-\textstyle{\frac{i}{4}}\theta^2 [e^{i\varphi}J_- -e^{-i\varphi}J_+,
 e^{i\varphi}J_- +e^{-i\varphi}J_+]\nn
&=i\theta^2 J_3,
\ena
and 
\bea
({\rm ad}_{-B})^2 (\p_\varphi B)
&=[-\textstyle{\frac{1}{2}}\theta (e^{i\varphi}J_- -e^{-i\varphi}J_+),i\theta^2 J_3]\nn
&=-\textstyle{\frac{i}{2}}\theta^3 [e^{i\varphi}J_- -e^{-i\varphi}J_+, J_3]\nn
&=-\textstyle{\frac{i}{2}}\theta^3 (e^{i\varphi}J_- +e^{-i\varphi}J_+)\nn
&=-\theta^2 \p_\varphi B.
\ena
Hence, the sum over even order terms in \eqref{dexp} is
\bea
\sum_{l:{\rm even}} \frac{1}{(l+1)!}({\rm ad}_{-B})^l (\p_\varphi B)
&=\p_\varphi B \left(1+\textstyle{\frac{1}{3!}}(-\theta^2)+ \textstyle{\frac{1}{5!}}\theta^4+\cdots\right)\nn
&=\textstyle{\frac{i}{2}}(e^{i\varphi}J_- +e^{-i\varphi}J_+)
\left(\theta -\textstyle{\frac{1}{3!}}\theta^3+ \textstyle{\frac{1}{5!}}\theta^5-\cdots\right)\nn
&=\textstyle{\frac{i}{2}}(e^{i\varphi}J_- +e^{-i\varphi}J_+)\sin\theta,
\ena
while the sum over odd order terms in \eqref{dexp} is
\bea
\sum_{l:{\rm odd}} \frac{1}{(l+1)!}({\rm ad}_{-B})^l (\p_\varphi B)
&=i\theta^2 J_3 \left(\textstyle{\frac{1}{2!}}+ \textstyle{\frac{1}{4!}}(-\theta^2)+ \textstyle{\frac{1}{6!}}\theta^4+\cdots\right)\nn
&=iJ_3 \left(\textstyle{\frac{1}{2!}}\theta^2 - \textstyle{\frac{1}{4!}}\theta^4+ \textstyle{\frac{1}{6!}}\theta^6+\cdots\right)\nn
&=iJ_3 (1-\cos\theta).
\ena
Combining them, we obtain
\bea
R^\dagger_N(\Omega) \p_\varphi R_N(\Omega )
&= d\exp_{-B}(\p_\varphi B) \nn
&=i(1-\cos\theta)J_3 +\textstyle{\frac{i}{2}}\sin\theta (e^{i\varphi}J_- +e^{-i\varphi}J_+),
\label{key relation for RdR2}
\ena
which is the same as \eqref{RNid3}.

The gauge potential in ${\cal U}_N$ is calculated as
\bea
iA_{N\theta} 
&=\Bra{\ell,+}R^\dagger_N(\Omega) \p_\theta R_N(\Omega ) \Ket{\ell,+}\nn
&=\Bra{\ell,+}\textstyle{\frac{1}{2}}(e^{i\varphi}J_- -e^{-i\varphi}J_+)\Ket{\ell,+}
=0,
\ena
from \eqref{key relation for RdR1} and 
\bea
iA_{N\varphi} 
&=\Bra{\ell,+}R^\dagger_N(\Omega) \p_\varphi R_N(\Omega )\Ket{\ell,+}\nn
&=i(1-\cos\theta)\Bra{\ell,+}J_3 \Ket{\ell,+} 
=i(1-\cos\theta) (\ell+\textstyle{\frac{1}{2}}) 
={\textstyle \frac{i}{2}}k(1-\cos\theta),
\ena
from \eqref{key relation for RdR2}, where we have used $k=2\ell+1$.
Then, we find
\bea
A_N=-i \Bra{\ell,+}R^\dagger_N(\Omega) d R_N(\Omega ) \Ket{\ell,+}
&={\textstyle \frac{1}{2}}k(1-\cos\theta) d\varphi.
\ena

For the metric, we need to evaluate
\bea
&C_{Nij}(\Omega) =\Bra{\ell,+}\p_i R^\dagger_N(\Omega ) \p_j R_N(\Omega )\Ket{\ell,+},
\ena
in addition to the gauge potential.
By using \eqref{RNid1} - \eqref{RNid4}, we obtain
\bea
C_{N\theta\theta}
&=\textstyle{\frac{1}{2}}(\ell+{\textstyle \frac{1}{2}}),\nn
C_{N\varphi\theta}
&=-\textstyle{\frac{i}{2}}(\ell+{\textstyle \frac{1}{2}})\sin\theta ,\nn
C_{N\theta\varphi}
&=\textstyle{\frac{i}{2}}(\ell+{\textstyle \frac{1}{2}})\sin\theta,\nn
C_{N\varphi\varphi}
&=(\ell+{\textstyle \frac{1}{2}})^2 (1-\cos\theta)^2 
+\textstyle{\frac{1}{2}}(\ell+{\textstyle \frac{1}{2}}) \sin^2\theta.
\ena
These are shown as follows: 
\bea
C_{N\theta\theta}
&=\Bra{\ell,+}\p_\theta R^\dagger_N(\Omega ) \p_\theta R_N(\Omega )\Ket{\ell,+}\nn
&=-\textstyle{\frac{1}{4}}\Bra{\ell,+}\left(e^{i\varphi}J_- -e^{-i\varphi}J_+\right)^2 
\Ket{\ell,+}\nn
&=\textstyle{\frac{1}{4}}\Bra{\ell,+}J_+ J_- \Ket{\ell,+}\nn
&=\textstyle{\frac{1}{4}}(2\ell +1),
\ena
where
\bea
J_+ J_- \Ket{\ell,+}
=J_+\left(\sqrt{2\ell}\Ket{\ell-1,+}+\Ket{\ell,-}\right)
=(2\ell +1) \Ket{\ell,+},
\ena
has been used.
\begin{align}
C_{N\varphi\theta}
&=\Bra{\ell,+}\p_\varphi R^\dagger_N(\Omega ) \p_\theta R_N(\Omega )\Ket{\ell,+}\nn
&=-\textstyle{\frac{1}{2}}\Bra{\ell,+}
\left[
i(1-\cos\theta)J_3 +\textstyle{\frac{i}{2}}\sin\theta (e^{i\varphi}J_- +e^{-i\varphi}J_+)
\right]
\left(e^{i\varphi}J_- -e^{-i\varphi}J_+\right)
\Ket{\ell,+}\nn
&=-\textstyle{\frac{1}{2}}\Bra{\ell,+}
\textstyle{\frac{i}{2}}\sin\theta J_+ J_-
\Ket{\ell,+}\nn
&=-\textstyle{\frac{i}{4}}\sin\theta (2\ell +1). \\
C_{N\theta\varphi}
&=\Bra{\ell,+}\p_\theta R^\dagger_N(\Omega ) \p_\varphi R_N(\Omega )\Ket{\ell,+}\nn
&=-\textstyle{\frac{1}{2}}\Bra{\ell,+}
\left(e^{i\varphi}J_- -e^{-i\varphi}J_+\right)
\left[
i(1-\cos\theta)J_3 +\textstyle{\frac{i}{2}}\sin\theta (e^{i\varphi}J_- +e^{-i\varphi}J_+)
\right]
\Ket{\ell,+}\nn
&=\textstyle{\frac{1}{2}}\Bra{\ell,+}
\textstyle{\frac{i}{2}}\sin\theta J_+ J_-
\Ket{\ell,+}\nn
&=\textstyle{\frac{i}{4}}\sin\theta (2\ell +1). \\
C_{N\varphi\varphi}
&=\Bra{\ell,+}\p_\varphi R^\dagger_N(\Omega ) \p_\varphi R_N(\Omega )\Ket{\ell,+} \nn
&=-\Bra{\ell,+}\left[
i(1-\cos\theta)J_3 +\textstyle{\frac{i}{2}}\sin\theta (e^{i\varphi}J_- +e^{-i\varphi}J_+)
\right]^2 \Ket{\ell,+}\nn
&=\Bra{\ell,+}\left[
(1-\cos\theta)^2(J_3)^2 +\textstyle{\frac{1}{4}}\sin^2\theta J_+ J_- \right] \Ket{\ell,+}\nn
&=\Bra{\ell,+}\left[(1-\cos\theta)^2 (\ell+{\textstyle \frac{1}{2}})^2 
+\textstyle{\frac{1}{4}}\sin^2\theta (2\ell+1)\right] \Ket{\ell,+}\nn
&=(\ell+{\textstyle \frac{1}{2}})^2 (1-\cos\theta)^2 
+\textstyle{\frac{1}{2}}(\ell+{\textstyle \frac{1}{2}}) \sin^2\theta.
\end{align}
By using these, the components in the metric $g_{ij}={\rm Re}(C_{ij}-A_i A_j)$ are found as 
\bea
&g_{\theta\theta}
=\textstyle{\frac{1}{2}}(\ell+{\textstyle \frac{1}{2}}),\nn
&g_{\theta\varphi}=g_{\varphi\theta}=0,\nn
&g_{\varphi\varphi}
=\textstyle{\frac{1}{2}}(\ell+{\textstyle \frac{1}{2}}) \sin^2\theta.
\ena

\paragraph{In the open set ${\cal U}_S$}
For the gauge potential $A_S$ in ${\cal U}_S$, 
we use the relation $R_S(\Omega )=R_N(\Omega )e^{-2i\varphi J_3}$ to write
\bea
R^\dagger_S(\Omega) d R_S(\Omega )
&=e^{2i\varphi J_3}R^\dagger_N(\Omega) d \left( R_N(\Omega )e^{-2i\varphi J_3}\right)\nn
&=e^{2i\varphi J_3}\left( R^\dagger_N(\Omega) d R_N(\Omega )\right) e^{-2i\varphi J_3}
+e^{2i\varphi J_3}d e^{-2i\varphi J_3}.
\label{shortcut}
\ena
This expression is valid only for ${\cal U}_{NS}$, but once we obtain $A_S$, it should be 
continued smoothly to the south pole.
Then, it is straightforward to show
\bea
A_S
&=-i \Bra{\ell,+}e^{2i\varphi J_3}\left( R^\dagger_N(\Omega) d R_N(\Omega )\right) e^{-2i\varphi J_3}\Ket{\ell,+}
-i \Bra{\ell,+}(-2i d\varphi J_3) \Ket{\ell,+}\nn
&=A_N -2 \Bra{\ell,+}J_3 \Ket{\ell,+}d\varphi \nn
&=A_N -2 (\ell+\textstyle{\frac{1}{2}}) d\varphi \nn
&=A_N -k d\varphi \nn
&=-{\textstyle \frac{1}{2}}k(1+\cos\theta) d\varphi.
\ena
For the metric, using \eqref{shortcut} again, we have the relations,
\bea
&\p_\theta R_S(\Omega ) 
=(\p_\theta R_N(\Omega )) e^{-2i\varphi J_3},\quad
\p_\varphi R_S(\Omega ) 
=(\p_\varphi R_N(\Omega ) -2i R_N(\Omega ) J_3) e^{-2i\varphi J_3}.
\ena
Thus, in order to obtain $C_{Sij}$, 
it is sufficient to evaluate the difference from $C_{Nij}$.
Apparently $C_{S\theta\theta}=C_{N\theta\theta}$, and 
\bea
C_{S\theta\varphi}
&=C_{N\theta\varphi}-2i \Bra{\ell,+}\p_\theta R_N^\dagger R_N J_3\Ket{\ell,+}
=C_{N\theta\varphi},\nn
C_{S\varphi\theta}
&=C_{N\varphi\theta}+2i \Bra{\ell,+}J_3 R_N^\dagger \p_\theta  R_N \Ket{\ell,+}
=C_{N\varphi\theta},\nn
C_{S\varphi\varphi}
&=C_{N\varphi\varphi}+\Bra{\ell,+}\left[2i(J_3R_N^\dagger\p_\varphi R_N
-\p_\varphi R_N^\dagger R_N J_3)+4(J_3)^2\right]\Ket{\ell,+} \nn
&=C_{N\varphi\varphi}+4(\ell+{\textstyle \frac{1}{2}})^2\left[-(1-\cos\theta)+1\right]
=C_{N\varphi\varphi}-2kA_{N\varphi}+k^2.
\ena
Combining these with $A_{S\theta}=0$ and $A_{S\varphi}=A_{N\varphi}-k$, 
we obtain the same metric as in ${\cal U}_N$. 
In particular, the difference in the $g_{\varphi\varphi}$ component vanish,
due to the cancellation of contributions from $C$ and $A$.


\providecommand{\href}[2]{#2}\begingroup\raggedright\endgroup

\end{document}